\documentclass[aps,prd,onecolumn,preprintnumbers,floatfix,showpacs,
nofootinbib,superscriptaddress,notitlepage]{revtex4-1}

\usepackage{amsmath}           %
\usepackage{amssymb}           %
\usepackage{color}             %
\usepackage{bbm}               %
\usepackage{braket}            %
\usepackage{xspace}            %
\usepackage{mciteplus}         %
\usepackage{mathtools}         %
\usepackage{graphicx}          %
\usepackage{slashed}           %
\usepackage{enumerate}         %
\usepackage[export]{adjustbox} %
\usepackage[english]{babel}    %
\usepackage[utf8]{inputenc}    %
\usepackage{hyperref}          %

\clearpage{}

\renewcommand{\d}{\mathbf{d}}
\renewcommand{\k}{\mathbf{k}}

\renewcommand{\P}{\mathbf{P}}

\newcommand{\0}{\mathbf{0}}
\newcommand{\1}{\mathbf{1}}
\newcommand{\2}{\mathbf{2}}

\newcommand{\n}{\mathbf{n}}
\newcommand{\m}{\mathbf{m}}

\newcommand{\Zbb}{\mathbbm{Z}}

\newcommand{\Cc}{\mathcal{C}}

\newcommand{\Fc}{\mathcal{F}}

\newcommand{\Jc}{\mathcal{J}}
\newcommand{\Kc}{\mathcal{K}}

\newcommand{\Mc}{\mathcal{M}}

\newcommand{\Oc}{\mathcal{O}}

\newcommand{\Rc}{\mathcal{R}}

\newcommand{\Wc}{\mathcal{W}}

\newcommand{\Zc}{\mathcal{Z}}

\DeclareMathOperator*{\SumInt}{%
   \mathchoice%
  {\ooalign{$\displaystyle\sum$\cr\hidewidth$\displaystyle\int$\hidewidth\cr}}
  {\ooalign{\raisebox{.14\height}{\scalebox{.7}{$\textstyle\sum$}}\cr\hidewidth$\textstyle\int$\hidewidth\cr}}
  {\ooalign{\raisebox{.2\height}{\scalebox{.6}{$\scriptstyle\sum$}}\cr$\scriptstyle\int$\cr}}
  {\ooalign{\raisebox{.2\height}{\scalebox{.6}{$\scriptstyle\sum$}}\cr$\scriptstyle\int$\cr}}
}

\newcommand{\bs}[1]{\boldsymbol{ #1 }}

\newcommand{\bh}[1]{\mathbf{\hat{ #1 }}}

\newcommand{\cf}{cf.\xspace}

\newcommand{\ie}{i.e.\xspace}

\newcommand{\df}{\textrm{df}}
\newcommand{\nn}{\nonumber}
\newcommand{\diff}{\textrm{d}}

\newcommand{\beq}{\begin{equation}}
\newcommand{\eeq}{\end{equation}}

\newcommand{\kev}{\ensuremath{{\mathrm{\,ke\kern -0.1em V}}}\xspace}
\newcommand{\mev}{\ensuremath{{\mathrm{\,Me\kern -0.1em V}}}\xspace}
\newcommand{\gev}{\ensuremath{{\mathrm{\,Ge\kern -0.1em V}}}\xspace}
\newcommand{\tev}{\ensuremath{{\mathrm{\,Te\kern -0.1em V}}}\xspace}

\usepackage{mfirstuc} %
\newcommand{\addReviewer}[2]{
  \expandafter\newcommand\csname #1\endcsname[1]{{\bf \color{#2} \capitalisewords{#1}:\,##1}}
  \expandafter\newcommand\csname #1cor\endcsname[2]{{\color{#2} \capitalisewords{#1}:\,\st{##1}{\bf ##2}}}
  \expandafter\newcommand\csname #1color\endcsname{#2}
}
\usepackage{soul,color}
\definecolor{chromeyellow}{rgb}{1.0, 0.65, 0.0}
\definecolor{DodgeBlue}{rgb}{0.118, 0.565,1.000}
\definecolor{asparagus}{rgb}{0.53, 0.66, 0.42}
\definecolor{cadmiumgreen}{rgb}{0.0, 0.42, 0.24}
\addReviewer{andrew}{cadmiumgreen}

\clearpage{}

\newcommand{\HHThree}[0]{Horz:2019rrn}
\newcommand{\BRSThree}[0]{Blanton:2019vdk}
\newcommand{\MaiThreeExcited}[0]{Mai:2019fba}
\newcommand{\RBCKpipi}[0]{Blum:2011ng,Boyle:2012ys,Blum:2015ywa,Bai:2015nea}
\hypersetup{ 
    pdfnewwindow=true,      %
    colorlinks=true,       %
    linkcolor=blue,          %
    citecolor=blue,        %
    filecolor=blue,      %
    urlcolor=blue        %
} 
\definecolor{jlab_red}{RGB}{192,39,45}
\definecolor{jlab_orange}{RGB}{249,102,0}
\definecolor{jlab_blue}{RGB}{47,122,121}
\definecolor{jlab_green}{RGB}{65,125,10}

\newcommand{\jlab}{Thomas Jefferson National Accelerator Facility, 
12000 Jefferson Avenue, Newport News, VA 23606, USA}
\newcommand{\odu}{Department of Physics, 
Old Dominion University, 
Norfolk, Virginia 23529, USA}
\newcommand{\cern}{Theoretical Physics Department, 
CERN, 1211 Geneva 23, Switzerland}

\begin{document}
\title{Consistency checks for two-body finite-volume matrix elements: \\
I. Conserved currents and bound states}
\author{Ra\'ul A. Brice\~no}
\email[e-mail: ]{rbriceno@jlab.org}
\affiliation{\jlab}
\affiliation{\odu}
\author{Maxwell T. Hansen}
\email[e-mail: ]{maxwell.hansen@cern.ch}
\affiliation{\cern}
\author{Andrew W. Jackura}
\email[e-mail: ]{ajackura@odu.edu}
\affiliation{\jlab}
\affiliation{\odu}
\preprint{JLAB-THY-19-3040}
\preprint{CERN-TH-2019-149}
\begin{abstract}
Recently, a framework has been developed to study form factors of two-hadron states probed by an external current. The method is based on relating finite-volume matrix elements, computed using numerical lattice QCD, to the corresponding infinite-volume observables. As the formalism is complicated, it is important to provide non-trivial checks on the final results and also to explore limiting cases in which more straightforward predications may be extracted. In this work we provide examples on both fronts. First, we show that, in the case of a conserved vector current, the formalism ensures that the finite-volume matrix element of the conserved charge is volume-independent and equal to the total charge of the two-particle state. Second, we study the implications for a two-particle bound state. We demonstrate that the infinite-volume limit reproduces the expected matrix element and derive the leading finite-volume corrections to this result for a scalar current. Finally, we provide numerical estimates for the expected size of volume effects in future lattice QCD calculations of the deuteron's scalar charge. We find that these effects completely dominate the infinite-volume result for realistic lattice volumes and that applying the present formalism, to analytically remove an infinite-series of leading volume corrections, is crucial to reliably extract the infinite-volume charge of the state.
\end{abstract}
\date{\today}
\maketitle

\section{Introduction}\label{sec:intro}

One of the overarching goals of modern-day nuclear physics is the characterization and fundamental understanding of the low-lying strongly-interacting spectrum. There is, by now, overwhelming evidence that the detailed properties of all low-lying states are governed by the dynamics of quark and gluon fields in the mathematical framework of quantum chromodynamics (QCD). But still, it remains a significant challenge to extract low-energy predictions from the underlying theory.

The vast majority of QCD states emerge as either bound states or resonances of multi-hadron configurations. An example is the deuteron, a shallow bound state of the isoscalar proton-neutron channel with a binding energy of $m_n + m_p - M_{d} \approx 2.2$~MeV. The deuteron has long been hypothesized to be a molecular state of the two nucleons~\cite{Weinberg:1962hj} and similar pictures have been proposed for a variety of other QCD states. (See Ref.~\cite{Guo:2017jvc} for a recent review.) However, in many cases a straightforward interpretation is unavailable. For example, the isoscalar $f_0(980)$ resonance couples strongly to $\pi\pi$ and $K\overline{K}$ states, and has been postulated to be both a tetraquark~\cite{Jaffe:1976ig} and a $K\overline{K}$ molecule~\cite{Weinstein:1990gu}.\footnote{Similar outstanding puzzles are present in the heavy quark sector; see Refs.~\cite{Lebed:2016hpi} and \cite{Brambilla:2019esw} for recent reviews.}

The challenge of resolving the inner structure of composite hadrons is twofold:
First, QCD is non-perturbative, so that systematic low-energy calculations are challenging. This has been addressed with substantial success using low-energy effective theories, methods based in amplitude analysis and numerical calculations using lattice QCD (LQCD). 
In contrast to the first two methods, LQCD has the unique advantage of relating the fundamental QCD lagrangian to low-energy predictions.
Second, composite states generally manifest as dynamical enhancements of multi-hadron scattering rates, meaning that the detailed observation depends on the production mechanism and decay channel of the resonance in question. This ambiguity is resolved, at least in principle, by recognizing that across all production and decay channels, a given resonance always leads to the same pole in an analytic continuation of scattering amplitudes to complex energies.

These two points have motivated the community to develop a systematic framework for extracting hadronic scattering amplitudes via LQCD. From the energy dependence of such amplitudes one can then quantitatively describe the bound and resonant states of the theory. In addition, by extracting transition amplitudes involving external currents, one can in principle access structural information of these states. 
In this work, we focus on an example in the latter class of the amplitudes, namely $\2 + \Jc \to \2$ transition amplitudes. We consider a method, first introduced in Refs.~\cite{Briceno:2015tza, Baroni:2018iau}, that allows one to determine such quantities from numerical LQCD.%

The primary formal challenge arises from the fact that LQCD calculations are necessarily performed in a finite Euclidean spacetime, where the definition of asymptotic states is obscured. 
One of the leading methods to overcome this issue is to derive and apply non-perturbative mappings between finite-volume energies and matrix elements (directly calculable via numerical LQCD) and infinite-volume scattering and transition amplitudes.%
\footnote{We point the reader to Refs.~\cite{Briceno:2017max} and \cite{Hansen:2019nir} for recent reviews detailing the progress of the field. See also Refs.~\cite{Hansen:2017mnd} and \cite{Bulava:2019kbi} for alternative methods to determine rates and amplitudes, which require significantly larger volumes as well as techniques to regulate the inverse Laplace transform.} 
This approach was first introduced by L\"uscher~\cite{Luscher:1986pf, Luscher:1991n1}, in seminal work relating the spectrum of two-particle states in a cubic volume with periodicty $L$, to the corresponding infinite-volume amplitudes. 
The idea has since been extended for arbitrary two-particle scattering~\cite{Rummukainen:1995vs, Kim:2005gf, He:2005ey, Davoudi:2011md, Hansen:2012tf, Briceno:2012yi, Briceno:2013lba, Briceno:2014oea, Romero-Lopez:2018zyy} and more recently to three particles~\cite{Hansen:2014eka,Hansen:2015zga,Mai:2017bge,Hammer:2017kms,Briceno:2017tce,Briceno:2018aml,Mai:2018djl,Briceno:2018mlh,Guo:2018zss,Blanton:2019igq}, with the latter currently limited to identical scalars (or pseudoscalars). 
The two-particle relations have made possible the determination of hadronic scattering amplitudes for a wide range of particle species~\cite{Dudek:2010ew,Beane:2011sc,Pelissier:2012pi,Dudek:2012xn,Liu:2012zya,Beane:2013br,Orginos:2015aya,Berkowitz:2015eaa,Lang:2015hza,Bulava:2016mks,Hu:2016shf,Alexandrou:2017mpi,Bali:2017pdv,Bali:2017pdv,Wagman:2017tmp,Andersen:2017una,Brett:2018jqw,Werner:2019hxc,Mai:2019pqr,Wilson:2019wfr}, including energies where multiple channels are kinematically open~\cite{Wilson:2014cna,Dudek:2014qha,Wilson:2015dqa,Dudek:2016cru,Briceno:2016mjc,Moir:2016srx,Briceno:2017qmb,Woss:2018irj,Woss:2019hse}. Most recently, the first LQCD calculations to constrain three-particle interactions using excited states were performed in Refs.~\cite{\HHThree,\BRSThree,\MaiThreeExcited}.

Electroweak interactions involving scattering states can also be accessed using LQCD, via a generalization of the methods described above. The seminal example in this sector is the work of Ref.~\cite{Lellouch:2000pv}, providing a formal method for determining the electroweak decay, $K \to \pi \pi$. More generally, in processes for which the effects of the electroweak sector can be treated perturbatively, the relevant amplitudes are given via the evaluation of QCD matrix elements, built from the appropriate currents together with multi-particle external states.
These ideas have been successfully developed for the case that either the initial or the final state couples strongly to two-particle scattering states~\cite{Lellouch:2000pv,Kim:2005gf,Christ:2005gi,Hansen:2012tf,Briceno:2014uqa,Briceno:2015csa,Agadjanov:2016fbd} and implemented in a number lattice QCD studies, most prominently to determine the $K \to \pi \pi$ decay amplitudes~\cite{\RBCKpipi} as well as the electromagnetic process $\pi \gamma^\star \to \pi \pi$~\cite{Feng:2014gba, Briceno:2015dca, Briceno:2016kkp, Feng:2018pdq, Alexandrou:2018jbt}. This progress motivates the consideration of more complicated electroweak transitions, in particular those with two hadrons in both the initial and final state.

As we discuss in detail in Sec.~\ref{sec:FFs}, $\2 + \Jc \to \2$ transition amplitudes allow one to extract elastic form factors of bound states and resonances, thereby providing direct information on the structure of these states and possibly resolving which models are most descriptive~\cite{Kaplan:1998sz,Chen:1999tn,Albaladejo:2012te}.
As compared to the transitions described in the preceding paragraph, the necessary formalism for these quantities is significantly more complicated~\cite{Bernard:2012bi, Briceno:2012yi, Briceno:2015tza, Baroni:2018iau}.\footnote{Reference \cite{Detmold:2004qn} was the first work to consider the coupling of an external current to finite-volume two-hadron states. In that publication, the authors consider the use of background fields in the context of a fixed-order expansion in a particular effective field theory, an alternative to the matrix elements of currents discussed here.} Building on previous work, in Ref.~\cite{Briceno:2015tza} two of us derived a model-independent relation between the corresponding finite-volume matrix elements, schematically denoted $\langle 2|\mathcal{J}|2\rangle_L$ (where $L$ indicates the side-length of the periodic cubic volume), and the $\2+\Jc\to\2$ transition amplitude, $\mathcal{W}$. In Ref.~\cite{Baroni:2018iau} we improved the method by simplifying technical details relating to the on-shell projection of the single-particle form factor and by using Lorentz covariant poles in the various finite-volume kinematic functions that arise.  We stress that the two approaches are equivalent and only differ in the exact definitions of unphysical, intermediate quantities. The results are derived to all orders in the perturbative expansion of a generic relativistic field theory, for any type of two-scalar channels, with generalizations to spin and coupled channels left to future work. Details of this formalism are reviewed in Sec.~\ref{sec:FV_fcns}.

The purpose of this work is to provide two non-trivial checks on the general relations of Refs.~\cite{Briceno:2015tza,Baroni:2018iau}, and also to demonstrate their predictive power even in simplified special cases. As a first check, in Sec.~\ref{sec:WTI} we demonstrate that the method is consistent with the consequences of the conserved vector current. In particular, the formalism predicts that the charge of a two-hadron finite-volume state is exactly equal to the sum of the constituent charges and independent of $L$. This relies on non-trivial relations between various $L$-dependent geometric functions, and a relation between the $\2 \to \2$ and $\2 + \Jc \to \2$ amplitudes that follows from the Ward-Takahashi identity. The second check, presented in Sec.~\ref{sec:FVME}, considers the analytic continuation of the formalism below two-particle threshold, for theories with an $S$-wave bound state. We show that the finite- and infinite-volume matrix elements coincide (once normalization factors are accounted for) up to term scaling as $e^{- \mathcal \kappa_{\text{B}} L}$, where $\kappa_{\text{B}}^2 = m^2 - M_{\text{B}}^2/4$ defines the binding momentum for two constituents of mass $m$, binding to a mass of $M_{\text{B}}$. 

Presently, LQCD calculations of light nuclei properties are being performed at unphysically heavy quark masses, for which the binding momenta exceed their real-world values~\cite{Beane:2012vq,Yamazaki:2015asa,Berkowitz:2015eaa,Wagman:2017tmp,Francis:2018qch}. In addition the properties of states can be shifted, e.g.~the dineutron, in nature a virtual bound state, is found to be a standard bound state for $m_\pi \gtrsim 450 \mev$ \cite{Beane:2011iw,Yamazaki:2011nd,Yamazaki:2012hi,Beane:2013br,Orginos:2015aya}. The increased binding suppresses finite-volume effects and this has permitted exploratory calculations of matrix elements of these states~\cite{Beane:2014ora,Chang:2017eiq,Savage:2017prl,Winter:2017bfs,Tiburzi:2017iux}, in which volume effects are ignored.

As LQCD calculations of multi-nucleon systems move towards physical quark masses, the binding momenta of the nuclei decrease and it is well-known that finite-volume effects of the naively extracted states can become a dominant source of systematic uncertainty~\cite{Davoudi:2011md, Briceno:2013bda, Briceno:2013hya}. In the case of spectroscopy, an infinite series of $e^{-  \kappa_{\text{B}} L}$ corrections can be removed by applying the L\"uscher formalism,
as was done in~\cite{Francis:2018qch, Berkowitz:2015eaa} as well as in a wide variety of mesonic channels where bound states appear~\cite{Wilson:2014cna,Dudek:2014qha,Lang:2015hza,Briceno:2016mjc,Moir:2016srx,Bali:2017pdv}. The results of this work stress that it is important to pursue the same paradigm for matrix elements of loosely bound states, using the formalism of Refs.~\cite{Briceno:2015tza, Baroni:2018iau} to non-perturbatively remove binding-momentum-enhanced finite-volume artifacts. To illustrate this point, in Sec.~\ref{sec:FVME} we determine the leading $e^{- \kappa_{\text{B}} L}$ corrections and compare these to the full result, which holds up to $e^{- m L}$. Finally, in Sec.~\ref{sec:num_expectation} we present a numerical example, meant to model the deuteron at physical pion masses, and show that the full formalism is needed to reliably remove the $L$-dependence for box sizes in the region of $mL \approx 4 - 7$. Otherwise the $e^{- \kappa_{\text{B}} L}$ corrections can become comparable in size with the infinite-volume result and thereby dominate the systematic uncertainties.

Though largely addressed above, we close here with a brief summary of the remaining sections. After reviewing basic properties of the infinite-volume $\2 \to \2$ and $\2 + \Jc \to \2$ amplitudes in Sec.~\ref{sec:FFs}, in Sec.~\ref{sec:FV_fcns} we described the corresponding finite-volume formalism for each type of amplitude. Then, in a very compact Sec.~\ref{sec:WTI}, we demonstrate that the finite-volume $\2 + \Jc \to \2$ formalism gives the expected results for matrix elements of a conserved current. Section~\ref{sec:BS_FV} is dedicated to volume effects on a two-particle bound state, including a check that the $L \to \infty$ limit gives the required result, a calculation of the leading $\mathcal O(e^{- \kappa_{\text{B}} L})$ corrections, and a numerical exploration intended to guide future LQCD calculations of the deuteron's scalar charge. We briefly conclude in Sec.~\ref{sec:conclusion}. In addition, this article includes three appendices, providing proofs of various technical results used in the main text.

\section{Infinite-volume amplitudes and bound states}\label{sec:FFs}

In this section we review the definitions and key properties of the infinite-volume $\2\to\2$ and $\2+\Jc\to\2$ amplitudes, with particular attention to the expressions relevant for an $S$-wave bound state.
For simplicity, we focus on systems composed of two scalar particles, with degenerate mass $m$, distinguished by their charge with respect to an external current $\mathcal J^\mu$. One of the particles carries charge ${\mathrm Q_0}$, while the other is neutral. Here we have in mind a scalar analog of the proton-neutron system.

\subsection{$\2\to\2$ amplitudes and bound-state poles}\label{sec:2to2}

In a general Lorentz frame, the two-particle system has a total energy-momentum denoted by $P = (E,\P)$. Boosting to the center-of-momentum frame (CMF) we define $P^\star = (E^\star, \0)$, which is related to the Mandelstam variable $s$ and a generic $P$ by
\begin{equation}\label{eq:s_inv}
E^{\star\,2}  \equiv s \equiv P_\mu P^\mu   = E^2 - \P^2.
\end{equation}
Two-particle scattering is described by $s$, as well as the back-to-back momentum orientations of the initial and final states in the CMF: $\bh{\k}_{i}^{\star}$ and $\bh{\k}_{f}^{\star}$, respectively.\footnote{We denote CMF quantities with a $\star$ superscript throughout.} Using these coordinates we can introduce the scattering amplitude and its partial wave expansion%
\begin{equation}\label{eq:M_expand}
\Mc(s,\bh{\k}_f^{\star},\bh{\k}_i^{\star}) = 4\pi \sum_{\ell, m_{\ell}} Y_{\ell m_{\ell}}(\bh{\k}_f^{\star}) \, \Mc_{\ell} (s) \,Y_{\ell m_{\ell}}^{*}(\bh{\k}_i^{\star}) \,.
\end{equation}
We have used that total angular momentum, $\ell$, is conserved, and that the partial-wave amplitude is independent of the projection, $m_{\ell}$, both consequences of rotational symmetry. 
In the following we are interested in the case of a scalar bound state, appearing as a sub-threshold pole in $\Mc_{\ell = 0} (s)$. We therefore restrict attention to the $S$-wave ($\ell = 0$) amplitude and do not write the angular momentum index on the partial wave amplitude for the rest of this section. %

\begin{figure}[t!]
    \centering
    \includegraphics[ width=0.4\textwidth]{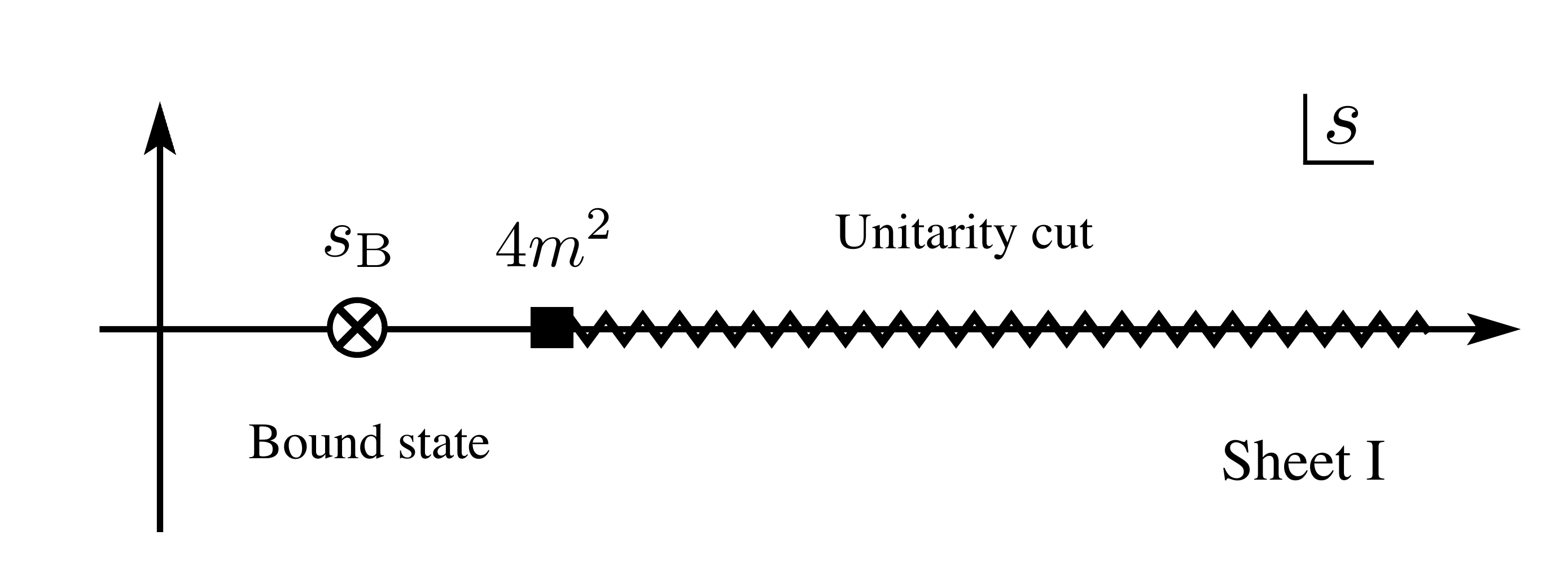}
    \caption{Analytic structure of $\Mc$ on the first Riemann sheet in the complex $s$-plane. Bound-state poles lie below two-particle threshold.}
    \label{fig:diag_bs_complex_plane}
\end{figure}

The elastic $\2\to\2$ scattering amplitude can be represented in terms of the K matrix, which is an on-shell representation that enforces S matrix unitarity explicitly below the inelasticity threshold~\cite{Martin:102663},
\begin{equation}\label{eq:Kmat_rep}
\Mc(s) = \Kc(s) \frac{1}{1 - i\rho(s) \, \Kc(s)}.
\end{equation}
Here, $\rho(s)$ is the two-body phase space, encoding the on-shell propagation of two particles. It is defined as
\begin{equation}\label{eq:rho}
\rho(s) = \frac{q^{\star}}{8\pi E^{\star}} =  \frac{1}{16\pi} \sqrt{1 - \frac{4m^2}{s}} ,
\end{equation}
where $q^{\star}$ is the relative momentum of the two particles in the CMF, $q^{\star} \equiv \sqrt{s/4 - m^2}$. This square root introduces a branch cut in the complex $s$ plane, illustrated in Fig.~\ref{fig:diag_bs_complex_plane}. Bound states are then defined as subthreshold poles on the first Riemann sheet, the sheet for which $\text{Im} \, q^\star > 0$.%
\footnote{Virtual bound states (e.g.~the dineutron) and resonances (e.g.~the $\rho$) arise on the second sheet, for which $\text{Im} \, q^\star < 0$.} 

The K matrix, $\Kc(s)$, is a real function describing all of the dynamics of the system. It can be written in terms of the scattering phase shift, $\delta(s)$, via
\begin{equation}\label{eq:Kmat_phase}
\Kc^{\,-1}(s) \equiv \rho(s) \cot\delta(s).
\end{equation}
Unlike $\mathcal M(s)$, the K matrix is an analytic function of $s$ in a domain around $s = (2m)^2$ set by the nearest left-hand cut. It follows that the effective range expansion
\begin{equation}\label{eq:ERE}
q^{\star} \cot\delta(s) = -\frac{1}{a} + \frac{1}{2} r  q^{\star 2} + \Oc(q^{\star 4}) \,,
\end{equation}
has a finite radius of convergence, and gives a useful description of $\mathcal K(s)$ and $\mathcal M(s)$ near threshold. The parameters $a$ and $r$ are called the scattering length and effective range, respectively. 

Using Eqs.~\eqref{eq:Kmat_rep}-\eqref{eq:Kmat_phase}, the condition for a bound state (a real sub-threshold pole on the first Riemann sheet) can be expressed as
\begin{equation}\label{eq:BS_pole}
q^{\star} \, \cot\delta(q^{\star})  \big\rvert_{q^{\star} = i\kappa_{\textrm{B}}} + \kappa_{\textrm{B}} = 0 \,,
\end{equation}
where $\kappa_{\textrm{B}}$ is the binding momentum, related to the pole position $s_{\text{B}}$ via $  s_{\textrm{B}}   = 4(m^2 - \kappa_{\textrm{B}}^2)$ where the mass of the bound state is given as $M_{\textrm{B}} = \sqrt{s_{\textrm{B}}}$.
Going beyond the pole, information about the nature and structure of the bound state is also contained in its coupling to two-particle scattering states, $g$, defined as the residue of the pole
\begin{equation}\label{eq:M_BS_pole}
\Mc(s) = \frac{(ig)^2}{s - s_{\textrm{B}} } \big [1 + \mathcal O(s - s_{\textrm{B}}) \big ]  \,.
\end{equation}
As we review in Sec.~\ref{eq:BS_FV_energies}, $g$ governs the prefactor of the bound state's leading finite-volume effects. 

\subsection{$\2+\Jc\to\2$ amplitudes and form factors}\label{sec:2Jto2}

We now turn to the less standard $\2+\Jc\to\2$ transition amplitude, defined via
\begin{equation}\label{eq:W_matrix}
\bra{P_f,\bh{\k}_f^{\star}, \text{out}} \Jc^{\mu} \ket{P_i, \bh{\k}_i^{\star}, \text{in}}_{\textrm{conn.}} \equiv \Wc^{\mu}(P_f,\bh{\k}_f^{\star}; P_i, \bh{\k}_i^{\star}) \,.
\end{equation}
Here the initial and final states have kinematics as in Sec.~\ref{sec:2to2} and the current, $\Jc^{\mu}$, is a local operator evaluated in position space at the origin. Since the current can inject energy and momentum, the initial and final states carry different total four-momenta, $P_i$ and $P_f$ respectively. It is also convenient to define the squared momentum transfer, $Q^2 \equiv - (P_f - P_i)^2$, where the overall minus is included so that $Q^2 > 0$ for spacelike $P_f - P_i$. 

The amplitude $\Wc^\mu$ can be defined for local currents with any Lorentz structure and in Sec.~\ref{sec:bs_scalarc} we also consider specific results for a scalar current. Here, for concreteness we focus on a conserved vector current $\Jc^{\mu}(x)$ satisfying
\begin{equation}
\partial_\mu \Jc^{\mu}(x) = 0 \,.
\end{equation}
\begin{figure}[t!]
    \centering
    \includegraphics[ width=0.60\textwidth]{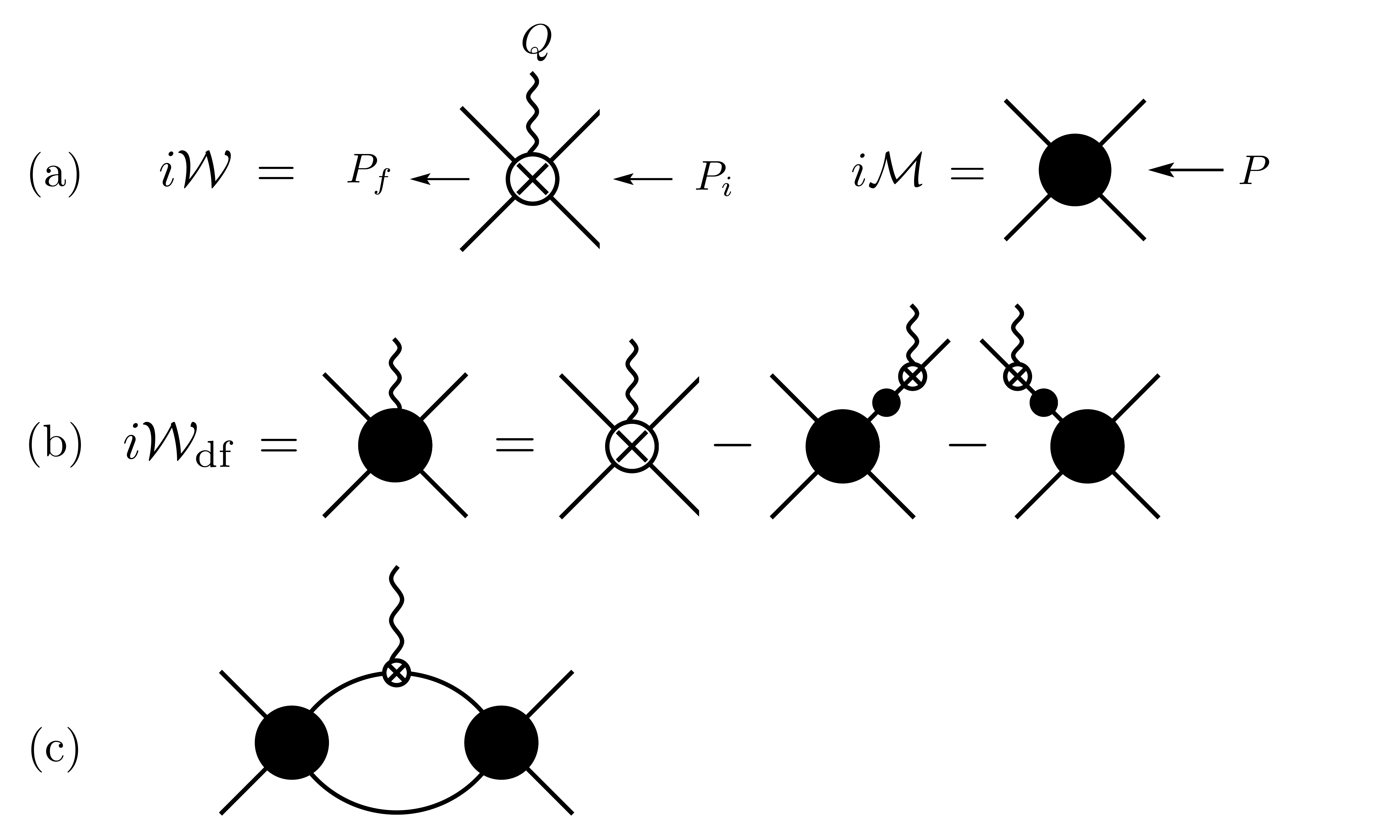}
    \caption{(a) Representation of (left) the $\2+\Jc\to\2$ amplitude with initial and final energy-momentum given by $P_i$ and $P_f$, respectively, leading to a current-induced momentum transfer of $Q^2 = -(P_f - P_i)^2$, as well as (right) the $\2 \to \2$ scattering amplitude with energy-momentum $P$. (b) Definition of $\Wc_{\df}$ where the long-range terms of a current probing an external leg are removed. (c) Triangle diagram leading to important singularities in $\Wc_{\df}$.}
    \label{fig:2Jto2_W_amp}
\end{figure}
Our first aim is to connect this amplitude to the bound-state form factor, defined via
\begin{equation}
\label{eq:FBdef}
\bra{P_f , {\text{B}}} \Jc^\mu \ket{P_i , {\text{B}}} = (P_f + P_i)^\mu F_{\text{B}}(Q^2)  \,,
\end{equation}
where $\ket{P_i , {\text{B}}}$ is the bound state, normalized as
$\braket{P_f , {\text{B}} | P_i , {\text{B}}}  =(2 \pi)^3  \, 2 \omega_{\P_i}   \delta^3(\textbf P_f - \textbf P_i)$ %
with energy $\omega_{\P} = \sqrt{ s_{\text{B}} + \textbf P^2 }$. 
Eq.~\eqref{eq:FBdef} is related to the $S$-wave projection of $\Wc^\mu$, analytically continued below threshold to the bound-state pole:
\begin{equation}\label{eq:W_BS}
\Wc^{\mu}(P_f,P_i) =  (P_i + P_f)^{\mu} \, F_{\textrm{B}}(Q^2)    \frac{i^2  (ig)^2}{( s_f - s_{\textrm{B} }) ( s_i - s_{\textrm{B} }) }  \big [1 + \mathcal O(s_{i,f} - s_{\text{B}}) \big ] \,,
\end{equation}
where $s_{i,f} = P_{i,f}^2$. We prove this result in Appendix \ref{app:Wpole}.

As discussed in some detail in Refs.~\cite{Briceno:2015tza, Baroni:2018iau}, the analytic structure of $\mathcal W^\mu$ is significantly more complicated than that of $\mathcal M$. One can identify three generic sources of non-analyticity in the $\2 + \Jc \to \2$ amplitude: (\emph{i}) Exactly as for $\mathcal M$, diagrams with on-shell two-particle intermediate states lead to factors of $\rho(s)$.  $\mathcal W^\mu$ thus exhibits the same branch cut and multi-sheet structure as $\mathcal M$. (\emph{ii}) Isolated poles arise due to the subtracted diagrams in Fig.~\ref{fig:2Jto2_W_amp}(b), in which the current is attached to an external leg.
(\emph{iii}) The triangle diagram shown in Fig.~\ref{fig:2Jto2_W_amp}(c) induces a new class of singularities, first described by Landau in Ref.~\cite{Landau:1959fi}.

To relate the $\2 + \Jc \to \2$ amplitude to a physical scattering rate, it is important to recall that the former arises from perturbatively expanding the weakly-interacting sector (encoded in $\mathcal J^\mu$) while keeping the strong dynamics non-perturbative. In fact, both the triangle singularities and the isolated poles can be understood as artifacts, resulting from truncating the expansion at a fixed order. Of course, the more standard single-particle form factors also arise from such an expansion, but happen to exhibit more straightforward analytic structure in the kinematic region considered. Indeed, as we will see below, analytic continuation to the bound-state pole removes all three of the non-analyticities we have identified.

As first explained in Ref.~\cite{Briceno:2015tza}, finite-volume matrix elements are more directly related to a subtracted amplitude, denoted $\Wc_{\df}$, from which the isolated poles, item (\emph{ii}) above, have been removed. Here the subscript `df' stands for `divergence free'. The definition, depicted also in Fig.~\ref{fig:2Jto2_W_amp}(b), reads
\begin{multline}\label{eq:Wdf_def}
 \Wc_{\df}^{\mu}(P_f,\bh{\k}_f^{\star}; P_i, \bh{\k}_i^{\star})  \equiv \Wc^{\mu}(P_f,\bh{\k}_f^{\star}; P_i, \bh{\k}_i^{\star})   \\ -  i \overline{\Mc}(P_f,k',k) \, \frac{i}{(P_f - k)^2 - m^2} \, w^{\mu}(P_f, P_i , k) - w^{\mu} (P_f, P_i , k') \, \frac{i}{(P_i - k')^2 - m^2} \, i\overline{\Mc}(P_i,k',k) \,,
\end{multline}
where $w^{\mu}$ is the single-particle matrix element,
\begin{align}\label{eq:single_w}
  w^{\mu} (P_f, P_i , k) & \equiv \bra{P_f - k , \text{Q}_0 } \Jc^{\mu}\ket{P_i - k, \text{Q}_0} \,, \\ %
& = (P_f+P_i - 2 k )^{\mu} \, f\big ( Q^2 \big)  \,,
\end{align}
and we have also introduced the corresponding form factor, $f$.
Here we adopt the convention that the charged particles carry momenta $P_i - k$ (incoming) and $P_f - k'$ (outgoing).\footnote{In these expressions we are dropping terms that depend on the form factor of the neutral particle. These can be readily included and only introduce slight technical complications in various equations. Such contributions are numerically suppressed compared to the charged particle's form factor, and vanish identically as $Q^2 \to 0$.
} The overline in $\overline {\mathcal M}$ denotes a slight modification to the definition of  $\Mc$ to account for the off-shell leg. This is described in Ref.~\cite{Baroni:2018iau} and, since the distinction is irrelevant for the $S$-wave amplitude, we do not discuss the issue further in this work.

The momentum directions within $\Wc_{\df}^{\mu}$ can be projected to definite angular momentum as done in Eq.~\eqref{eq:M_expand} for $\mathcal M$
\begin{equation}\label{eq:Wdf_pwe}
\Wc_{\df}^{\mu}(P_f,\bh{\k}_f^{\star}; P_i, \bh{\k}_i^{\star}) = 4\pi \sum_{\substack{ \ell_f,m_f \\ \ell_i,m_i} } Y_{\ell_f m_f}(\bh{\k}_f^{\star}) \, \Wc_{\df, \ell_f m_f, \ell_i m_i}^{\mu}(P_f,P_i) \,Y_{\ell_i m_i}^{*}(\bh{\k}_i^{\star}) \,.
\end{equation}
As with the scattering amplitude, for the remainder of this section we restrict attention to the $S$-wave component of $\Wc_{\df}^{\mu}$, i.e.~the component containing our bound state. Note that, in contrast to $\mathcal M$, $ \Wc_{\df}$ has off diagonal elements in angular momentum space, due to the angular momentum injected by the external current.
A crucial observation that will guide our later analysis is that the $S$-wave component of $\Wc_{\df}$
exactly satisfies Eq.~\eqref{eq:Wdf_BS} above, i.e.~$\Wc^{\mu}$ and $\Wc_{\df}^{\mu}$ have the same bound-state double-pole with the same residue:
\begin{equation}\label{eq:Wdf_BS}
\Wc_{\df}^{\mu}(P_f,P_i) =  (P_i + P_f)^{\mu} \, F_{\textrm{B}}(Q^2)    \frac{i^2  (ig)^2}{( s_f - s_{\textrm{B} }) ( s_i - s_{\textrm{B} }) }  \big [1 + \mathcal O(s_{i,f} - s_{\text{B}}) \big ] \,.
\end{equation}
This equivalence holds because the subtracted terms in Eq.~\eqref{eq:Wdf_def} only have a single pole, and thus cannot modify the leading divergence. 

The bound-state poles within $\Wc_{\df}^\mu$ motivate us to introduce a new object, $\Fc^{\mu}(P_f,P_i)$, given by
\begin{equation}\label{eq:Wdf_on}
\Wc_{\df}^{\mu}(P_f,P_i) = \Mc(s_f) \, \Fc^{\mu}(P_f,P_i) \, \Mc(s_i) \,. 
\end{equation}
The $S$-wave scattering amplitudes on each side remove the poles from $\Wc_{\df}^{\mu}(P_f,P_i)$, implying
\begin{align} \label{eq:FF_K}
\lim_{s_i, s_f \to s_{\text{B}}}  \Fc^{\mu}(P_f,P_i) &  = (P_f + P_i)^\mu \frac{F_{\textrm{B}}(Q^2)}{g^2} \,.
\end{align}
In addition, Eq.~\eqref{eq:Wdf_on} factorizes the $\rho(s)$ branch cuts from $\Wc_{\df}^{\mu}$ so that $\mathcal F^\mu$ does not contain this class of singularities.\footnote{We leave a detailed proof of this claim and an analysis of its consequences to future work~\cite{analytic}.}
Nonetheless, $\Fc^{\mu}$ is in general complex, due to the triangle diagram of Fig.~\ref{fig:2Jto2_W_amp}(c). This diagram can only contribute complexity (as well as non-analyticity) when at least one of the two-particle cuts goes on shell, i.e.~when such an intermediate state can physically propagate. In particular, for subthreshold energies, and thus for some domain around the bound-state energy, the triangle integral is real and analytic.

At this stage we have argued that each of the three non-analyticities listed above is irrelevant near the bound-state pole: First, the external-leg poles are removed in the conversion from $\Wc$ to $\Wc_{\df}$, second, the on-shell threshold cuts in $\rho(s)$ are removed in the relation between $\Wc_{\df}^{\mu}$ and $ \Fc^{\mu}$, and finally, the triangle singularity (still contained in $\Fc^\mu$) is avoided by the subthreshold continuation.

We close this section with an important property of $\Wc_{\df}^{\mu}$ that follows from the Ward-Takahashi identity. As we sketch in Appendix~\ref{app:WTI} and derive in detail in Ref.~\cite{analytic}, $\Wc_{\df}^{\mu}$ satisfies the following simple relation to the scattering amplitude
\begin{equation}\label{eq:WTI_W}
\Wc_{\df}^{\mu} (P) = {\mathrm Q_0} \frac{\partial}{\partial P_{\mu}} \Mc(s) = 2P^{\mu} {\mathrm Q_0} \frac{\partial}{\partial s}\Mc(s) \,.
\end{equation}
This identity is crucial for the analysis presented in Sec.~\ref{sec:Ltoinf}.

\section{Finite-volume formalism for two-particle systems}\label{sec:FV_fcns}

Before giving detailed expressions for the finite-volume effects on a two-body bound state, in this section
we briefly review the general formalism describing the finite-volume energies and matrix elements of two-particle systems. In the following, we work in a cubic, periodic volume of length $L$ with infinite temporal extent. The total momentum of the system in the finite-volume frame is allowed to take on any value consistent with the periodicity: $\P = 2\pi   \n / L$ with $\n \in \Zbb^{3}$.

\subsection{Finite-volume energies}\label{sec:FV_spectrum}

In the window of energies for which only two particles can propagate, the finite-volume spectrum is related to the infinite-volume partial-wave amplitudes, defined in Eq.~\eqref{eq:Kmat_rep}, via the L\"uscher quantization condition~\cite{Luscher:1991n1, Rummukainen:1995vs, Kim:2005gf}. 
Generally, the quantization condition is a determinant over angular momentum space. If we neglect waves higher than $\ell = 0$, however, it reduces to a simple algebraic relation
\begin{equation}\label{eq:Luscher}
\Mc^{-1}(s_n) = -F(P_n,L) + \Oc(e^{- m L}) \,,
\end{equation}
where $s_n = P_n^2 = E_n(L)^2 - \P^2$ corresponds to the eigenenergy of the $n$th finite-volume two-particle state. 
Here $F(P,L)$ is a known finite-volume function,
\begin{align}\label{eq:FV_F} 
F(P,L) & =   \bigg[  \frac{1}{L^{3}}  \SumInt_{\k} \bigg] \frac{1 }{2\omega_{\k} 2\omega_{\P\k} (E - \omega_{\k} - \omega_{\P\k} + i\epsilon)} \,, %
\end{align}
where $\omega_{\k} = \sqrt{m^2 + \k^2}$ and $\omega_{\P\k} = \sqrt{m^2 + (\P-\k)^2}$ are the on-shell energies of the two particles, and
\[
\bigg[  \frac{1}{L^{3}}  \SumInt_{\k} \bigg] \equiv \frac{1}{L^{3}} \sum_{\k \in (2\pi/L) \Zbb^{3} } - \int \frac{ \diff^{3}\k}{(2\pi)^{3} }.
\]
Equation~\eqref{eq:Luscher} holds up to corrections associated with higher partial waves and only for $s_n$ below the first inelastic threshold.

\subsection{Finite-volume matrix elements}\label{sec:FV_matrix}

Similarly, one can relate finite-volume matrix elements of two-particle systems to infinite-volume $\2 + \Jc \to \2$ transition amplitudes. 
Here, the relevant formalism was first derived in Ref.~\cite{Briceno:2015tza} using an all-orders perturbative expansion based in a generic relativistic effective field theory. 
Recently, in Ref.~\cite{Baroni:2018iau}, the formal approach was improved in two ways: First, by rearranging the separation of finite-volume effects, we were able to show that the extracted infinite-volume transition amplitudes are manifestly Lorentz covariant. Second, we re-organized the analysis so that single-particle matrix elements enter via standard form factors (rather than a non-standard spherical harmonic decomposition used in the first publication).
 While the two representations are formally equivalent, the work of Ref.~\cite{Baroni:2018iau} is expected to be significantly more convenient in numerical applications going forward. Of course, all expressions used here are taken from the improved approach.

\bigskip

Again, assuming all but the $\ell=0$ partial waves are negligible, the matrix elements of the vector current for two-particle states can be related to $\Wc_{\df}^{\mu}$, defined in Eq.~\eqref{eq:Wdf_on}, as follows
\begin{equation}\label{eq:BH_eqns}
L^{3} \bra{P_{n,f},L} \Jc^{\mu} \ket{P_{n,i},L} = \Wc_{L,\df}^{\mu}(P_{n,f},P_{n,i},L) \, \sqrt{ \Rc(P_{n,f},L) \Rc(P_{n,i},L) },
\end{equation}
where $P_{n,i} = (E_{n,i},\P_i)$ and $P_{n,f} = (E_{n,f},\P_f)$. Here $\Wc_{L,\df}^{\mu}$ is an $L$-dependent function related to $\Wc_{\df}^{\mu}$ in a manner detailed in the following paragraph. In addition, $\Rc$ is a generalization of the Lellouch-L\"uscher factor \cite{Lellouch:2000pv}, first introduced in Ref.~\cite{Briceno:2014uqa}
\begin{align}\label{eq:Rdef}
\Rc(P_n,L) 
& =  \left[ \frac{\partial}{\partial E} \left( F^{-1}(P,L) + \Mc(s) \right) \right]^{-1}_{E = E_n} \,, \\
& = \label{eq:Rdef2}
 -  \Mc^{-2}(s_n)   \left[ \frac{\partial}{\partial E} \left( F(P,L) + \Mc^{-1}(s)   \right) \right]^{-1}_{E = E_n} \,,
\end{align}
where we have given a second form that will be particularly useful for this work.
In general, $\Rc$ is a matrix over all two-particle degrees of freedom, but in the case considered it reduces to a simple derivative of the functions shown. 

Before defining $\Wc_{L,\df}^{\mu}(P_f,P_i,L)$, we need to introduce a second $L$-dependent kinematic function, $G^{\mu_1\cdots\mu_n}$, first introduced in Refs.~\cite{Briceno:2015tza,Baroni:2018iau}.
For the $\ell = 0$ truncation it takes the form
\begin{equation}\label{eq:FV_G}
G^{\mu_1 \cdots \mu_n} (P_f,P_i,L) = \bigg[  \frac{1}{L^{3}}  \SumInt_{\k} \bigg] \frac{ k^{\mu_1} \cdots k^{\mu_n} }{2\omega_{\k}  ((P_{f} - k)^2 - m^2 + i\epsilon) ((P_{i} - k)^2 - m^2 + i\epsilon) } \Big\rvert_{k^{0} = \omega_{\k}}.
\end{equation}
In this work we will specifically need the scalar and vector $G$-functions, denoted $G$ and $G^{\mu}$ respectively. These are defined by keeping zero or one factor, respectively, of $k^\mu$ in the numerator of the integrand.
With these in hand, $\Wc_{L,\df}^{\mu}$ can be defined via its relation to $\Wc^\mu_{\df}$ as follows:
\begin{align}
\label{eq:WLdf_vector}
\Wc^{\mu}_{L,\df}(P_{f},P_{i} , L) & = \Wc^{\mu}_{\df }(P_{f}, P_{i}) 
+    f(Q^2)  \Mc(s_f)  \Big [  (P_f + P_i)^{\mu}  G(P_{f},P_{i},L)  - 2   G^{\mu}(P_{f},P_{i},L)    \Big ]  \Mc (s_i) \, .
 \end{align}
Here $f(Q^2)$ is the form factor of the charged particle while the form factor of the neutral particle, which vanishes identically at $Q^2=0$, is assumed negligible for all values of momentum transfer. 

\section{Matrix elements of the conserved vector current}\label{sec:WTI}

Having introduced the general formalism, we proceed to perform the checks outlined in the introduction. The first check is to show that, for any finite-volume state, the matrix element with respect to the charge operator
\begin{equation}
\widehat{\mathrm{Q}} \equiv \int \diff^3 \textbf x \, \mathcal J^0(x) \,,
\end{equation}
is predicted by the formal mapping to be $L$-independent and equal to the charge of the state. 
To demonstrate this, we first introduce another expression for the Lellouch-L\"uscher factor.  
Evaluating the energy derivative of $F$ in Eq.~\eqref{eq:Rdef2}, one can show
\begin{align}\label{eq:R_Kmat}
\Rc(P_n(L),L)  &  =  \frac{1}{\Mc^2(s_n(L))} \left[ 
  -\frac{\partial}{\partial E} \Mc^{-1}(s)
+ 2 E \, G(P ,L) - 2G^{\mu=0}(P ,L)
\right]^{-1}_{P = P_n(L)} \,.
\end{align}
Here we have also adopted the shorthand $G(P ,L) \equiv G(P, P,L)$, i.e.~we do not repeat the total momentum argument when it is the same for the incoming and outgoing states.
Note that, in this subsection, we are considering not only the finite-volume bound state but also excited states. We do continue to restrict attention to the $S$-wave only.

Substituting this result into Eq.~\eqref{eq:BH_eqns}, and also taking the relation between $\Wc_{\df}^\mu$ and $\mathcal F^\mu$ [Eq.~\eqref{eq:Wdf_on}], we find
\begin{multline}\label{eq:fv_matrix_complicated}
L^3 \bra{P_{n,f},L} \Jc^{\mu} \ket{P_{n,i},L} = \\  \frac{   \Fc^{\mu}(P_f, P_i) + f(Q^2)  \big[ (P_i + P_f)^{\mu}G(P_f,P_i,L) - 2G^{\mu}(P_f,P_i,L) \big]   }{ \sqrt{\big [- {\partial}_{E_i} \Mc^{-1}(s_i)   + 2E_i  G(P_i,L) - 2G^{\mu=0}(P_i,L) \big ]  \big [-{\partial}_{E_f} \Mc^{-1}(s_f)   + 2E_f  G(P_f,L) - 2G^{\mu=0}(P_f,L) \big ] }}  \bigg \vert_{P_{i,f} = P_{i,f}(L)}
\,.
\end{multline}
This result will prove very powerful in the following derivations. 
To see the consequences of this for the charge operator we set $\mu=0$ in the vector current and also set the initial and final-states to coincide. This yields
\begin{equation}\label{eq:fv_matrix_simple}
  \bra{P_{n},L} \widehat {\mathrm{Q}}  \ket{P_{n},L} = \\  \frac{   \Fc^{0} (P) + f(0)  \big[ 2 E G(P ,L) - 2G^{0} (P ,L) \big]   }{   - {\partial}_{E } \Mc^{-1}(s )   + 2E   G(P ,L) - 2G^{0} (P ,L) }  \bigg \vert_{P  = P_n(L)}
\,,
\end{equation}
where we have used the $\textbf x$-independence of the matrix element to replace $L^3 \mathcal J^0\!(0) \to \widehat {\mathrm{Q}} $ and have defined $\Fc^{0}(P) \equiv \Fc^{0}(P,P)$ as a convenient shorthand for systems with identical initial and final momenta.

This can be further simplified via the identity 
\begin{equation}\label{eq:F_constraint}
\Fc^{0}(P)  =\frac{ {\mathrm Q_0} }{ \Mc^{2}(s)} \frac{\partial}{\partial E} \Mc(s)  = -   {\mathrm Q_0} \frac{\partial}{\partial E}  \Mc^{-1}(s)  \,,
\end{equation}
which immediately follows from Eqs.~\eqref{eq:Wdf_on} and \eqref{eq:WTI_W}. Substituting this into the numerator of Eq.~\eqref{eq:fv_matrix_simple} and also using $f(0) = \mathrm Q_0$, we recover a very satisfying cancellation of all terms to deduce
\begin{equation}
  \bra{P_{n},L} \widehat {\mathrm{Q}}  \ket{P_{n},L} = \mathrm Q_0 \,,
\end{equation}
as expected.
This is a highly non-trivial verification that the general $\2+\Jc\to\2$ finite-volume formalism is consistent the consequences of current conservation. The derivation relies on two unexpected identities: First, the fact that the energy-derivative of $F(P,L)$ can be expressed using the $G$-functions, as shown in Eq.~\eqref{eq:R_Kmat}, and second, that the Ward-Takhashi identity relates $\mathcal F^{0}(P)$ to the scattering amplitude, Eq.~\eqref{eq:F_constraint}.

\section{Bound state in a finite volume}\label{sec:BS_FV}

 We now turn to the implications of the general formalism for bound-state matrix elements in a finite volume.

\subsection{Volume effects on the energies}\label{eq:BS_FV_energies}

 We start by reviewing results for finite-volume effects in the energy level, $E^{\textbf P}_{\textrm{B}}(L)$, defined to coincide with the moving bound state in the infinite-volume limit,
\begin{equation} 
\lim_{L \to \infty}   E^{\textbf P}_{\textrm{B}}(L) = E^{\textbf P}_{\textrm{B}} \equiv \sqrt{ M_{\textrm{B}}^2 + \P^2 } \,.
\end{equation}
Boosting these energies to the rest frame, we also define
\begin{equation}
s^{\textbf P}_{\textrm{B}}(L) \equiv E^{\textbf P}_{\textrm{B}}(L)^2 - \textbf P^2 \equiv s_{\textrm{B}} + \delta s^{\textbf P}_{\textrm{B}}(L) \,,
\end{equation}
with $s_{\textrm{B}} = \lim_{L \to \infty}   s^{\textbf P}_{\textrm{B}}(L) = M_{\textrm{B}}^2$. Note that the finite-volume energies depend on $\textbf P$, even after boosting back to the rest frame. In the following, we give expressions for the volume-induced shift, $\delta s^{\textbf P}_{\textrm{B}}(L)$, for two values of total momentum. This represents a small subset of the more general expressions derived in Ref.~\cite{Davoudi:2011md}.\footnote{Related results for bound states can be found in Refs.~\cite{Beane:2003da, Bour:2011ef, Briceno:2013bda, Briceno:2013hya}.}

The quantization condition, Eq.~\eqref{eq:Luscher}, is satisfied only at the finite-volume energies, e.g.~at $P_{\textrm{B}}(L)\equiv( E^{\textbf P}_{\textrm{B}}(L), \textbf P )$. We are thus strictly interested in $F(P,L)$ only when it is evaluated at these points. However, taking $\delta s_{\textrm{B}}$ as a small parameter, we note
\begin{equation}
F (P_{\textrm{B}}(L), L) =   F (P_{\textrm{B}} , L)  +  \Oc(\delta s_{\textrm{B}}) \,,
\end{equation}
where $P_{\textrm{B}} \equiv ( E^{\textbf{P}}_{\textrm{B}}, \textbf P )$ is the infinite-volume bound-state momentum in a moving frame.

As is discussed in detail in Ref.~\cite{Davoudi:2011md} and reviewed in Appendix~\ref{app:cont}, the subthreshold $L$-dependence of the $F$-function is governed by the binding momentum: $ \kappa_{\text{B}}^2 \equiv m^2    - M_B^2/4$.
In particular from Eqs.~\eqref{eq:Fc1_def} and \eqref{eq:cn100_asymp} we find
\begin{equation}\label{eq:FV_F_kappa}
F(P_{\textrm{B}},L) = -\frac{1}{8\pi   M_{\textrm{B}}   } \sum_{\m\ne\0}  \, e^{iL \m \cdot \P / 2} \,  \frac{ e^{- \kappa_{\text{B}} L \lvert\m'\rvert}}{ L \lvert\m'\rvert},
\end{equation}
where
\begin{equation}
\m' \equiv \m + (\gamma - 1) \frac{\m \cdot \P}{\lvert \P \rvert^2} \P \,,
\label{eq:mp}
\end{equation}
and $\gamma = E^{\textbf{P}}_{\textrm{B}}/M_{\textrm{B}}$. This result is to be combined with the inverse scattering amplitude, also evaluated at $s^{\textbf P}_{\textrm{B}}(L)$, but then expanded in powers of $\delta s_{\textrm{B}}$ to yield
\begin{align}
\Mc^{-1}(s^{\textbf P}_{\textrm{B}}(L)) &=  \delta s_{\textrm{B}}\frac{\partial}{\partial s} \Mc^{-1}(s) \Big\rvert_{s = s_{\textrm{B}}}  + \mathcal O(\delta s_{\textrm{B}}^2) \,, \\[3pt]
&=  - \delta s_{\textrm{B}} /g^2  + \mathcal O(\delta s_{\textrm{B}}^2) \,,
\label{eq:Mexp}
\end{align}
where we have used $\Mc^{-1}(s_{\textrm{B}}) = 0$.
Combining Eqs.~\eqref{eq:FV_F_kappa} and \eqref{eq:Mexp} then yields the elegant result
\begin{align}  \label{eq:delta_sB}
   \delta s^{\textbf P}_{\textrm{B}}(L)  &=  g^2 F(P_{\textrm{B}},L)  + \mathcal O(e^{- 2 \kappa_{\text{B}} L})\, ,
\end{align}
which shows that the leading shift to the finite-volume bound state is given directly by the $F$-function, evaluated at the infinite-volume bound-state energy.

To close this section we think it useful to unpack Eq.~\eqref{eq:delta_sB} for a two specific cases. First, in the case of vanishing momentum in the finite-volume frame, the three universal orders are given by
\begin{equation}
  \delta s^{[000]}_{\textrm{B}}  =  -\frac{6 g^2}{8\pi M_{\textrm{B}} L  } \bigg [ e^{-\kappa_{\textrm{B}} L } +  \sqrt{2}      e^{- \sqrt{2} \kappa_{\textrm{B}} L }  + \frac{4}{3 \sqrt{3}}  e^{- \sqrt{3} \kappa_{\textrm{B}} L  } \bigg ]  + \mathcal O(e^{- 2 \kappa_{\text{B}} L}) \,.
 \end{equation}
 At $\mathcal O(e^{- 2 \kappa_{\text{B}} L})$ higher derivatives of the inverse amplitude enter, requiring information beyond the coupling, $g$.
 
For nonzero momenta, the expressions are complicated by the relation between $\textbf m'$ and $\textbf m$, and by the volume dependence entering $\gamma$ through $E^{\textbf{P}}_B = \sqrt{ M_{\text{B}}^2 + (2 \pi/L)^2 \n^2 }$. Useful results can be reached, however, by expanding in all $L$ dependence. Performing such an expansion, and neglecting terms scaling as $ e^{-\kappa_{\textrm{B}} L }/L^2$ and $ e^{- \sqrt{2} \kappa_{\textrm{B}} L }$, we find
\begin{align}
  \delta s^{[00n]}_{\textrm{B}}  & =  -\frac{  g^2 [4 + 2 \cos(n \pi)] e^{-\kappa_{\textrm{B}} L }}{8\pi M_{\textrm{B}} L  } 
  \bigg [1     -  \frac{ n \cos (  n \pi)}{4 + 2 \cos (  n \pi)}  \frac{4 \pi ^2 \kappa_{\textrm{B}}}{    M_{\textrm{B}}^2 L}   \bigg ]  \,.
\end{align}
 
 To compare these results to those in Ref.~\cite{Davoudi:2011md} we note that, in the earlier work, the authors introduce an $L$-dependent binding momentum, defined via $\kappa_{\text{B}}(L)^2 = m^2 - s_{\text{B}}(L)/4$. Then the finite-volume shift, $\delta \kappa_{\text{B}}(L) \equiv \kappa_{\text{B}}(L) - \kappa_{\text{B}}$ satisfies the relation
\begin{equation}
\delta s^{\textbf P}_{\textrm{B}}(L) = -8 \kappa_{\textrm{B}} \,\delta\kappa^{\textbf P}_{\textrm{B}}(L) + \mathcal O(\delta s_{\textrm{B}}^2 )\,.
\end{equation}
Combining this with the relation between the coupling and the scattering phase
\begin{equation}
\frac{1}{g^2} =   \frac{  1  }{64 \pi \kappa_{\text{B}}  M_{\text{B}}}   \bigg (1 - 2 \kappa_{\text{B}} \frac{\diff}{\diff q^{\star\,2}}  q^{\star} \cot \delta(q^{\star} )   \bigg )_{\!\! s = s_{\text{B}}} \,,
\end{equation}
 yields Eq.~(9) of Ref.~\cite{Davoudi:2011md}.

In closing we comment that, due to the reduction of rotational symmetry, higher partial waves do induce finite-volume corrections to the scalar bound state and corresponding matrix elements. In particular, for $\P = 0$, $\ell=0$ mixes with $\ell =4, 6, \ldots\,$, as can be seen by the fact that the corresponding off-diagonal components of $F$ are nonzero. These additional angular-momentum contributions are, in fact, not volume-suppressed relative to the $S$-wave contributions, but are suppressed by powers of the binding momentum in units of the scattering-length analogs appearing in higher-partial waves. For example the $\ell=4$ phase shift satisfies an expansion analogous to Eq.~\eqref{eq:ERE}
\begin{equation}
q^{\star} \cot\delta_{\ell=4}(s) =  \frac{M_4^9}{q^{\star 8}} + \mathcal O(q^{\star -6}) \,,
\end{equation}
where $M_4$ has units of energy. In the case of zero spatial momentum in the finite-volume frame, one can show that the first non $S$-wave contribution to $s_{\text{B}}^{[000]}(L)$ is suppressed relative to the leading shift by a factor of $ \kappa_{\text{B}}^{8} \, M_{\text{B}} \, / \, M_4^9$.

Having reproduced the known expansion for the binding energy~\cite{Davoudi:2011md}, we now turn to the finite-$L$ corrections of the bound-state matrix element. 

\subsection{Volume effects on the matrix elements\label{sec:FVME}}

\subsubsection{Matrix elements in the $L\to\infty$ limit{\label{sec:Ltoinf}}}

We begin by confirming that, in the $ L \to\infty$ limit, the finite-volume bound-state matrix element (as described by the general $\2 + \Jc \to \2$ formalism) coincides with its infinite-volume counterpart. Here it is important to stress that the various quantities we consider have a well-defined $L \to \infty$ limit, only because we are considering them at sub-threshold kinematics and thus away from a set of finite-volume poles that becomes arbitrarily dense. 

We begin with Eq.~\eqref{eq:WLdf_vector}, the relation between $\Wc_{L,\df}^{\mu}$ and $\Wc_{\df}^{\mu}$. For $L \to \infty$, these two quantities coincide because the $G$ function defining their difference vanishes. This is the case because the sum within $G$ (\cf~Eq.~\eqref{eq:FV_G}) is transformed to an integral in the limit and is exactly canceled by the second, subtracted integral. Equation~\eqref{eq:BH_eqns} thus becomes
\begin{equation}\label{eq:BH_eqns_Linf}
\lim_{L \to \infty} L^3 \bra{P_{\text{B},f},L} \Jc^{\mu} \ket{P_{\text{B},i},L} = \lim_{L \to \infty}
 \Wc_{\df}^{\mu}(P_{\text{B},f}(L),P_{\text{B},i}(L) ) \,  \sqrt{ \Rc(P_{\text{B},f}(L),L) \,  \Rc(P_{\text{B},i}(L),L) } \,.
\end{equation}
The next step is to expand $\mathcal R$, evaluated at the finite-volume bound-state energy, about large $L$. Using the form given by Eq.~(\ref{eq:Rdef2}), one readily finds
\begin{align}
  \Rc(P_{\text{B}}(L),L) & = -  \Mc^{-2}(s_{\textrm{B}}^{\P}(L)) \bigg [ \frac{2 E_{\text{B}}}{g^2} + \Oc(e^{- \kappa_{\text{B}} L})\bigg ]^{-1} \,,   \\
  & = - \frac{ \big ( s_{\textrm{B}}^{\P}(L) - s_{\textrm{B}} \big )^{2} }{2E_{\textrm{B}} g^2} \Big [ 1  +  \mathcal O(e^{- \kappa_{\text{B}} L})\Big ] \,.
  \label{eq:Rexp}
\end{align}

We are now in position to evaluate the limit. The only subtlety is that a double-zero, arising from the Lellouch-L\"uscher factors, is exactly canceled by the double pole in $\Wc_{\df}^\mu$. Substituting Eqs.~\eqref{eq:Wdf_BS} and \eqref{eq:Rexp} into Eq.~\eqref{eq:BH_eqns_Linf}, we reach
\begin{equation}\label{eq:BS_deep_matrix}
\lim_{L \to \infty} 2 \sqrt{E_{\text{B},i} E_{\text{B},f} } L^3 \bra{P_{\text{B},f},L} \Jc^{\mu} \ket{P_{\text{B},i},L}  = \bra{P_f , {\text{B}}} \Jc^\mu \ket{P_i , {\text{B}}}   =  (P_{\text{B},i} + P_{\text{B},f})^{\mu} F_{\textrm{B}}(Q^2)    \,.
\end{equation}
This is exactly the desired result, with the extra factors on the left-hand side accounting for the different normalization conventions of finite- and infinite-volume states.

In this derivation we did not make reference to the Lorentz structure of the current, only to the fact that the $\2 + \Jc \to \2$ amplitude, $\mathcal W$, must have a double pole structure associated with the initial and final bound states. As a result, in general the formalism fulfills the expectation that for an arbitrary current $\Jc_{\mu_1\ldots \mu_n}$
\begin{equation}\label{eq:BS_deep_matrix_anycurrent}
\lim_{L \to \infty} 2 \sqrt{E_{\text{B},i} E_{\text{B},f} } L^3 \bra{P_{\text{B},f},L} \Jc^{\mu_1\ldots \mu_n} \ket{P_{\text{B},i},L} = \bra{P_f , {\text{B}}}  \Jc^{\mu_1\ldots \mu_n} \ket{P_i , {\text{B}}}   \,.%
\end{equation}

\subsubsection{Large $L$ expansion of the bound-state matrix element}\label{sec:bs_scalarc}

As shown in Sec.~\ref{sec:WTI}, the conserved vector current leads to volume-independent matrix elements at zero momentum transfer. Thus, to reach an interesting large-$L$ expansion, in this section we turn to a scalar current $\mathcal J$ and define
\begin{align}\label{eq:BH_scalar}
g^{\P}_{S,\text{B}}(L) & \equiv 2 E_B(L) L^3 \bra{P_{\text{B}},L} \Jc \ket{P_{\text{B}},L}  \,,
\end{align}
where the subscript indicates that this matrix element defines the scalar charge of the bound state. 
The infinite-volume bound-state scalar charge is recovered in the $L\to \infty$ limit, \ie $g_{S,\text{B}}  \equiv \lim_{L \to \infty} g^\P_{S,\text{B}}(L)$.
In direct analogy to Eq.~\eqref{eq:fv_matrix_simple} above, we observe
  \begin{equation}\label{eq:fv_matrix_simple_scalar}
  g^{\P}_{S,\text{B}}(L) =   \frac{  \mathcal F(s) + g_{S}  G(P,L)   }{ - \partial_s \Mc^{-1}(s)  +  G(P,L) - G^{ 0}  (P,L)/E  }  \Big\rvert_{P = P_{\text{B}}(L)} \,.
\end{equation}
Here $\Fc(s) \equiv  \Mc^{-2}(s) \Wc_{\df}(P,P)$ with $\Wc_{\df}$ given by Eq.~\eqref{eq:Wdf_def}, in which the vector current is replaced by a scalar. 
Note that the numerator includes only the scalar $G$-function, reflecting the scalar current considered. However, the denominator remains identical to the vector case since the Lellouch-L\"uscher factors are independent of the current. 
We have also introduced $g_{S}$ as the scalar charge of the single-particle state, $g_S  \equiv f(0)$, where $f$ is the single particle form-factor
\begin{align}
f(Q^2) & \equiv \bra{ P_f,  g_S } \Jc \ket{ P_i,  g_S } \,.
\end{align}
As above, we take the coupling of the current to the other constituent particle to be negligible.

With these ingredients in hand it is straightforward to expand Eq.~\eqref{eq:fv_matrix_simple_scalar} about $L \to \infty$ to reach
\begin{equation}\label{eq:BH_scalar}
\frac{  g^{\P}_{S,\text{B}} (L)} {  g_{S,\text{B}}}
 =        1 + \delta s^{\P}_{\text{B}}(L) \frac{\partial}{\partial s}  \bigg [  \frac{F_{\text{B}}(s)}{g_{S,\text{B}} }  + g^2  \frac{\partial}{\partial s} \Mc^{-1}(s)  \bigg ]+  \frac{g^2 (  g_{S}  - g_{S,\text{B}}   )}{g_{S,\text{B}}}     G(P_{\text{B}},L)           + \frac{ g^2 G^{ 0}  (P_{\text{B}},L)}{ E_{\text{B}}}  + \mathcal O(e^{- \sqrt{2}  \kappa_{\text{B}} L}) \,.
\end{equation}
We note that a great deal of structural information enters the leading finite-volume correction. The $\delta s_{\text{B}}(L)$-dependent term is the correction induced from the energy shift and is thus proportional to energy derivatives of both the inverse amplitude and the $\2 + \Jc \to \2$ transition amplitude (entering via $\mathcal F(s)$). The second term in Eq.~\eqref{eq:BH_scalar} arises due to a mismatch between the scalar charge of the bound-state and the summed charges of its constituents. 
The final term in Eq.~\eqref{eq:BH_scalar} is a direct consequence of the triangle diagram, Fig.~\ref{fig:2Jto2_W_amp}(c).

We close this section with a final, more explicit result for the leading-volume correction in the case where the CMF and finite-volume frames coincide, i.e.~$\textbf P = \0$. Substituting the leading result for $\delta s_{\text{B}}^{[000]}$, and results from Appendix \ref{app:cont} for the $G$-functions, one finds
\begin{equation}\label{eq:BH_scalarv2}
\frac{  g^{[000]}_{S,\text{B}} (L)} {  g_{S,\text{B}}}
 =        1   + g^2 \frac{3  e^{-\kappa_{\textrm{B}} L } }{32 \pi M_{\text{B}} \kappa_{\text{B}} } \bigg [1+ 2  \frac{   g_{S}  - g_{S,\text{B}}    }{g_{S,\text{B}}}   +  \frac{4   \kappa_{\text{B}}}{  M_{\text{B}}^2 L}  - \frac{ 8 \kappa_{\text{B}}}{   L  }  \frac{\partial}{\partial s}  \bigg (  \frac{F_{\text{B}}(s)}{g_{S,\text{B}} }  + g^2  \frac{\partial}{\partial s} \Mc^{-1}(s)  \bigg )_{\! \! s = s_{\text{B}}}        \bigg ]
   + \mathcal O(e^{- \sqrt{2}  \kappa_{\text{B}} L}) \,.
\end{equation}
The leading $1$ in the square brackets arrises from the triangle diagram, Fig.~\ref{fig:2Jto2_W_amp}(c), and will be the dominant finite-volume effect provided $|g_S-g_{S,B}|\leq |g_{S,B}|/2$.

\subsection{Numerical expectations for finite-volume dependence}\label{sec:num_expectation}

\begin{figure}[t!]
    \centering
    \includegraphics[ width=0.6\textwidth]{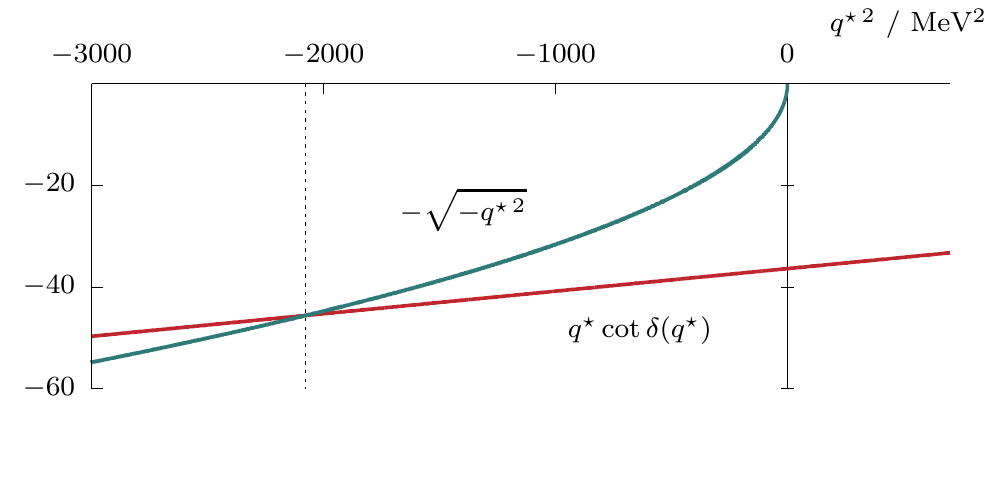}
    \caption{Plot of $q^{\star}\cot\delta(q^{\star})$ and $-\sqrt{-q^{\star2}}$ as a function of $q^{\star2}$, in units of $\mev^2$, for the effective range expansion, Eq.~\eqref{eq:ERE}, using the scattering length and effective range for the $pn$-system in $^{3}S_{1}$. The vertical dashed line indicates the deuteron, with binding momentum $\kappa_{\textrm{B}} \sim 45.58$ MeV.}
    \label{fig:fig-qcot_deut}
\end{figure}
\begin{figure*}[t!]
    \centering
    \includegraphics[ width=0.92\textwidth]{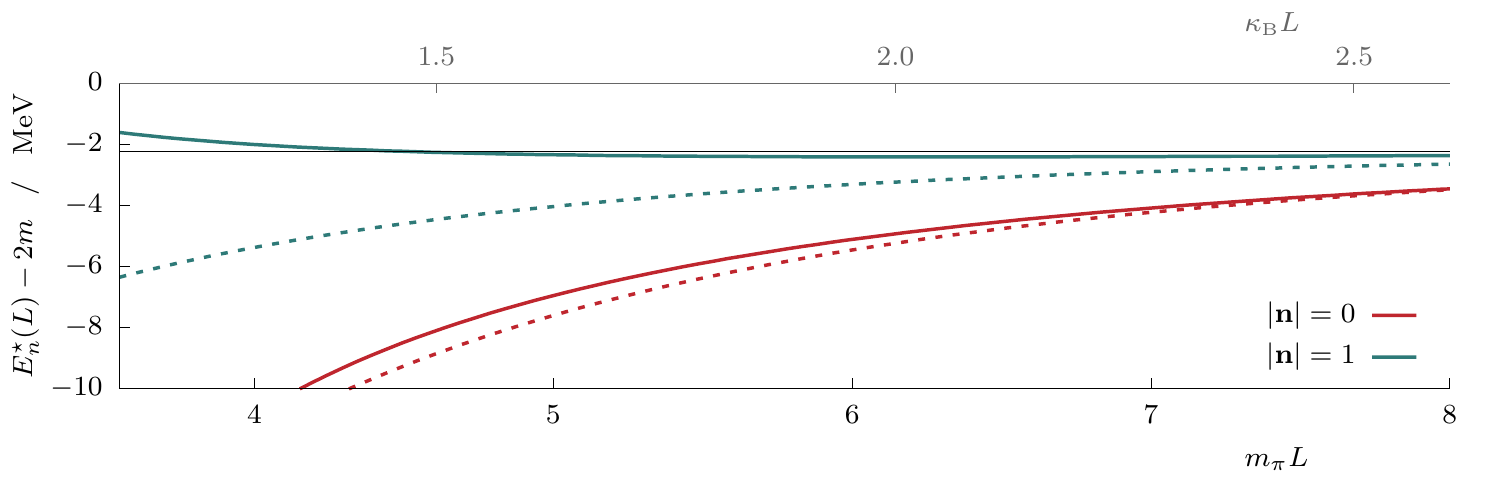}\label{fig:fig-energies}
    \put(-490,80){\colorbox{white}{(a)}}

    \includegraphics[ width=0.92\textwidth]{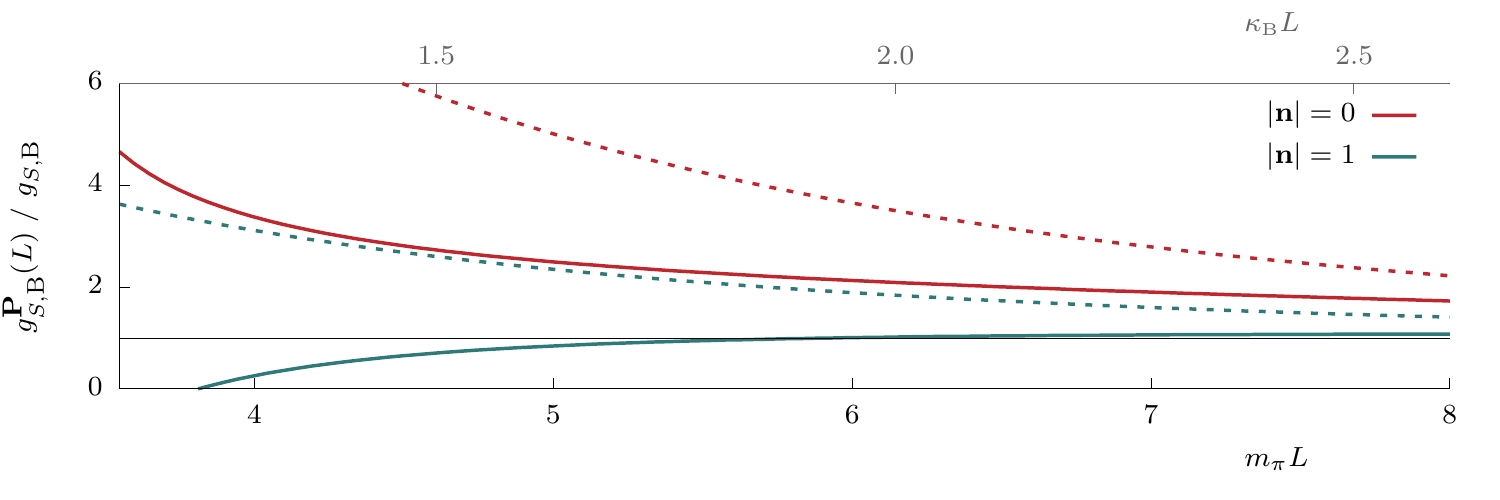}\label{fig:fig-matrix}
    \put(-490,80){\colorbox{white}{(b)}}
    \caption{(a) Finite-volume energy spectrum as a function of $m_{\pi}L$ for $pn$-scattering parameters as explained in the text. The two colors indicate a system at rest in the finite-volume frame (red) and a system that is boosted with one unit of momentum,  $\P = 2\pi  {\textbf n} / L$ and $\vert \textbf n \vert =1$ (blue-green). The solid lines show the prediction of the L\"uscher quantization condition, which holds up to $e^{- m_\pi L}$. The dashed lines show a prediction based on the leading $\Oc(e^{-\kappa_{\textrm{B}} L} )$ term in the large-$L$ expansion. All four curves asymptote to the horizontal line at $M_{\text{B}} -2m \sim -2.21$ MeV, the infinite-volume binding energy. %
    (b) Ratio of the finite-volume bound-state matrix element, $g_{S,\textrm{B}}^{\P}(L) = 2E_{\textrm{B}}(L) L^{3}\bra{P_{\textrm{B}},L}\Jc\ket{P_{\textrm{B}},L}$, to the infinite-volume scalar charge, $g_{S,\textrm{B}}$, as a function of $m_{\pi}L$. The solid curves show the prediction of the full formalism and the dashed lines show the leading term in the large-$L$ expansion, given by Eq.~\eqref{eq:BH_scalarv2} and its moving-frame analog. All four curves asymptote to $1$.
}
    \label{fig:fig-results}
\end{figure*}

In this section, we use the full $\2 + \Jc \to \2$ formalism to explore the finite-volume corrections to a bound-state matrix element in an example with scattering parameters chosen to mimic the deuteron.
As above we consider the simplest case of a scalar current and an $S$-wave bound state and examine the finite-volume corrections given by Eqs.~\eqref{eq:fv_matrix_simple_scalar} and ~\eqref{eq:BH_scalar}. For the $\2 \to \2$ scattering amplitude, we use use the phenomenological values for the $pn$ scattering length ($a = 5.425$ fm) and effective range ($r = 1.749$ fm) to describe the scattering amplitude and spectrum.  With the nucleon mass at $m = 934$ \mev\footnote{We work here with isospin symmetry, approximating $m_p = m_n$.}, the deuteron bound-state pole lies at $\sqrt{s_{\textrm{B}}} = 1875.63$ \mev with a coupling of $g = 5370.7$ \mev. This corresponds to a binding momentum of $\kappa_{\textrm{B}} = 45.58$ \mev, with a binding energy $-2.21$ \mev. The effective range expansion is shown in Fig.~\ref{fig:fig-qcot_deut} as a function of $q^{\star\,2}$ and the location of the bound state is indicated.

We make two assumptions to simplify the numerical exercise: First, we assume that the infinite-volume bound-state form factor is constant, \ie $F_{\textrm{B}}(s) = g_{S,\textrm{B}}$. 
As seen in the fourth term of the brackets in Eq.~\eqref{eq:BH_scalarv2}, this contribution is suppressed by $1/L$, thus it is reasonable that this approximation will not strongly alter the prediction.
Second, we assume that difference $g_{S} - g_{S,\textrm{B}}$ is numerically small and set $g_{S} = g_{S,\textrm{B}}$. 

Within this set-up one can numerically evaluate Eqs.~\eqref{eq:fv_matrix_simple_scalar} and \eqref{eq:BH_scalar} and compare the results. The first step is to determine the finite-volume bound-state energy, using the effective-range description of the $pn$ scattering amplitude in the quantization condition, Eq.~\eqref{eq:Luscher}.
Figure~\ref{fig:fig-results}(a) shows the bound-state energy as a function of $L$ for both $\lvert\d\rvert = 0$ and 1. 
The solid lines represents the full solution obtained from Eqs.~\eqref{eq:Kmat_rep} and \eqref{eq:Luscher} with the $pn$-scattering parameters. 
The dashed lines correspond to the leading-order approximation using Eq.~\eqref{eq:delta_sB} for the same momenta. 
These results reproduce those of Ref. \cite{Davoudi:2011md}, and we see that for lattice calculations performed at $m_{\pi}L \sim 4$, deviations between the exact and approximated forms are significant. %

Turning to the two-particle matrix elements, Fig.~\ref{fig:fig-results}(b) shows the ratio of the finite-volume bound-state matrix element to the infinite-volume scalar charge, $g_{S,\textrm{B}}^{\P}(L) \, / \, g_{S,\textrm{B}}$. 
Solid lines represent the full solution, using Eq.~\eqref{eq:fv_matrix_simple_scalar} evaluated at the finite-volume energy, and the dashed lines give the leading-order shift of Eq.~\eqref{eq:BH_scalar}. 
Again, significant deviations arise between the full prediction, the leading-order expansion, and the infinite-volume result.
This illustrates that, to reliably extract infinite-volume matrix elements of shallow bound states like the deuteron, it is highly beneficial to use the full formalism which removes an infinite series of terms scaling as powers of $e^{- \kappa_{\text{B}}L}$.   In the present example, only at $m_{\pi }L \sim 8$ do corrections scale to the percent level.

\section{Conclusion}\label{sec:conclusion}

In this work, we have provided strong consistency checks on, and also explored various consequences of, the formalism derived in Refs.~\cite{Briceno:2015tza, Baroni:2018iau}, which gives a relation between finite-volume matrix elements, schematically denoted  $\langle 2 \vert \Jc \vert 2 \rangle_L$, and the corresponding infinite-volume $\2+\Jc\to\2$ amplitudes. 

First, in the case of the conserved vector current, we have shown that resulting prediction for the two-particle matrix element of the charge operator,  $\langle 2 \vert \widehat {\textrm{Q}} \vert 2 \rangle_L$, behaves as expected. Specifically, the matrix element is $L$-independent and equal to the sum of the constituent charges. Though it is clear that this relation must hold,  the way it arises in the mapping is highly non-trivial, relying on an identity relating various $L$-dependent geometric functions [Eq.~\eqref{eq:R_Kmat}] as well as a relation between the $\2\to\2$ and $\2+\Jc\to\2$ amplitudes that follows from the Ward-Takahashi identity [Eq.~\eqref{eq:WTI_W}].

Second, for a generic local current, we have demonstrated that the mapping of Refs.~\cite{Briceno:2015tza, Baroni:2018iau}, reproduces the expected behavior in the case of an $S$-wave two-particle bound state. By analytically continuing the formal relations below threshold to the bound-state pole, we have confirmed that the finite- and infinite-volume matrix elements are equal up to volume corrections scaling as $e^{- \kappa_{\text{B}} L}$, where $\kappa_{\text{B}}$ is the binding-momentum of the state. This is an expected extension of the well-known result for the $L$ dependence of the bound-state  energy.

These two checks give confidence that our admittedly complicated formalism correctly describes two-particle finite-volume states and is ready to be implemented in a LQCD calculation, with the first application likely being the $(\pi \pi)_{\text{I=1}} + \Jc_\mu \to  (\pi \pi)_{\text{I=1}}$ transition amplitude, allowing one to extract the electromagnetic form factors of the $\rho$. 

As an additional example of the utility of the general approach, we have determined the full functional form of the leading, $\mathcal O(e^{- \kappa_{\text{B}} L})$ correction to the bound-state matrix element of a local scalar current. The result, Eq.~\eqref{eq:BH_scalarv2}, shows that the coefficient of the leading exponential depends on the bound state's coupling to the two-particle asymptotic state, the scalar charges of both the bound state and its constituents, and also on derivatives of both the $\2 \to \2$ scattering amplitude and the bound-state form factor.

While the structure of this relatively simple prediction is instructive, we stress that in practice it is more useful to use the general relation of Refs.~\cite{Briceno:2015tza, Baroni:2018iau} to extract the $\2+\Jc\to\2$ over a range of energies, including in a neighborhood around the bound-state pole. Doing so removes an infinite series of terms scaling as powers of $e^{- \kappa_{\text{B}} L}$ and, for shallow bound-states, allows one to control an otherwise dominant source of systematic uncertainty. To stress this point, as a final exercise, we have presented numerical comparisons of the leading $e^{- \kappa_{\text{B}} L}$ correction with the full finite-volume shift, for a toy set-up mimicking a LQCD calculation of the deuteron's scalar charge. For physical pion masses and volumes in the range $m_{ {\pi}}L \sim 4$ to $7$, we find that the finite-$L$ correction will dominate the infinite-volume charge and that removing only the leading exponential also does not give a reliable extraction. Thus, we conclude that the full method must be used to gain a reliable result for the form factors of shallow bound states as well as resonances.

This work makes use of identities that will be presented in a companion article that outlines, in detail, the analytic structure of the generalized form factors considered here~\cite{analytic}. 
An additional check is underway to reproduce analytic expressions for the $1/L$ expansion presented in Ref.~\cite{Detmold:2014fpa}, for the threshold-state matrix element of a scalar current in a weakly-coupled system.

\section{Acknowledgements}
We thank Alessandro Baroni, Felipe Ortega-Gama, and Akaki Rusetsky for useful discussions.
RAB is supported in part by USDOE grant No. DE-AC05-06OR23177, 
under which Jefferson Science Associates, LLC, manages and operates Jefferson Lab.
RAB also acknowledges support from the USDOE Early Career award, contract de-sc0019229. 

\appendix

\section{Proof of Eq.~\eqref{eq:W_BS}\label{app:Wpole}}

In this appendix we demonstrate that, in theories with a two-particle bound state, the $\2 + \Jc \to \2$ amplitude satisfies Eq.~\eqref{eq:W_BS}, repeated here for convenience
\begin{equation} 
\Wc^{\mu}(P_f,P_i) =  (P_i + P_f)^{\mu} \, F_{\textrm{B}}(Q^2)    \frac{i^2  (ig)^2}{( s_f - s_{\textrm{B} }) ( s_i - s_{\textrm{B} }) }  \big [1 + \mathcal O(s_{i,f} - s_{\text{B}}) \big ] \,.
\end{equation}
To show this, it is necessary to return to the matrix element definition of the amplitude, $\mathcal W^\mu$, given in Eq.~\eqref{eq:W_matrix}. Inserting a complete set of states on either side of the current, $\mathcal J^\mu$, we reach
\begin{equation}\label{eq:W_compstates}
\int \! \!  \frac{\diff^3 \textbf P''}{(2 \pi)^3 2 E_{\text{B}}^{\textbf{P}''} } \,   \int \! \! \frac{\diff^3 \textbf P'}{(2 \pi)^3 2 E_{\text{B}}^{\textbf{P}'} } \, \langle {P_f,\bh{\k}_f^{\star}, \text{out}}  \vert {P'', \text{B}} \rangle \langle  {P'', \text{B}}   \vert \,  \Jc^{\mu}  \,     \vert   P', \text{B} \rangle \langle  P', \text{B}    \vert   {P_i, \bh{\k}_i^{\star}, \text{in}}   \rangle  = \Wc^{\mu}(P_f,\bh{\k}_f^{\star}; P_i, \bh{\k}_i^{\star}) + \cdots \,,
\end{equation}
where we have kept only the bound-state sector of the Fock space, as this will be sufficient to identify the pole that we are after. 

Three additional subtleties arise here: (1) To properly implement the normalization of the bound-state, 
\begin{equation}
\langle  P', \text{B}   \vert  P, \text{B}  \rangle \equiv (2 \pi)^3 2 E_{\text{B}}^{\textbf{P}} \delta^3(\textbf{P}' - \textbf{P})  \,,
\end{equation}
we must integrate over all spatial momenta with the standard Lorentz-invariant factor as shown. (2) Since the spectral decomposition can only be performed on the full matrix element we have dropped the ``conn'' subscript that appears in Eq.~\eqref{eq:W_matrix}. To preserve the definition we have included the ellipsis on the right-hand side, which is understood to represent all disconnected contributions. These will, however, play no role, since they do not contain the bound-state pole. (3) The expression we are after requires the analytic continuation of $P_f$ and $P_i$ to the sub-threshold region. This is subtle at the level of Fock states, and is more easily understood by rewriting the result in terms of operators projected to definite momentum. This, in turn, reveals that the time-ordering of the operators must be carefully treated, as we explain in more detail below.

The next step is to substitute
\begin{align}
 \langle {P_f,\bh{\k}_f^{\star}, \text{out}}  \vert {P'', \text{B}} \rangle & \equiv (2 \pi)^4 \delta^4(P_f - P''_{\text{B}} )  \,  i g \,, \\
  \langle  P', \text{B}    \vert   {P_i, \bh{\k}_i^{\star}, \text{in}}   \rangle & \equiv (2 \pi)^4 \delta^4(P_i - P'_{\text{B}})   \, i g \,.
\end{align}
Here the four-dimensional delta function arises in direct analog to the standard relation between T matrix and scattering amplitude and leads to the definition of the bound-state coupling $g$. 
Using the spatial delta functions to evaluate the integrals in Eq.~\eqref{eq:W_compstates}, we reach
\begin{equation}
 \int_{-\infty}^\infty \diff x_0'' \, e^{i x_0'' (E_f - E_{\text{B}}^{\textbf P''}) } \,  \int_{-\infty}^\infty \diff x_0' \, e^{-i x_0' (E_i - E_{\text{B}}^{\textbf P'})}  \, \frac{i g }{{2 E_{\text{B}}^{\textbf{P}''} } } \langle  {P'', \text{B}}   \vert \,  \Jc^{\mu}   \,  \vert P', \text{B} \rangle \, \frac{  i g }{{2 E_{\text{B}}^{\textbf{P}'} }} = \Wc^{\mu}(P_f,\bh{\k}_f^{\star}; P_i, \bh{\k}_i^{\star}) + \cdots \,,
\end{equation}
where it is understood that one must set $\textbf P'' \to \textbf P_f$ and $\textbf P'' \to \textbf P_i$.
Here we have also written the remaining temporal delta functions as integrals over time. 

Introducing the integrals over $x_0''$ and $x_0'$ allows us to address the subtlety mentioned as point (3) above. Studying the correlation functions reveals that the above expression does not correctly treat all time orderings. For the present case, this is resolved by restricting the integral over $x_0''$ from $0$ to $\infty$ and similarly that over $x_0'$ from $-\infty$ to $0$. Doing so, and also including the $i \epsilon$ prescription required to project the external states in the correlator to the vacuum, one can evaluate both integrals to reach
\begin{equation}
\frac{i(ig) }{2 E_{\text{B}}^{\textbf{P}''} (E_f - E_{\text{B}}^{\textbf P''}) }   \,  \frac{i (ig) }{2 E_{\text{B}}^{\textbf{P}'} (E_i - E_{\text{B}}^{\textbf P'}) }    \, \langle  {P'', \text{B}}   \vert \,  \Jc^{\mu}   \,  \vert P', \text{B} \rangle \,   = \Wc^{\mu}(P_f,\bh{\k}_f^{\star}; P_i, \bh{\k}_i^{\star}) + \cdots \,.
\end{equation}
This is the result that we had aimed to prove. Up to the $\mathcal O((s_{i,f} - s_{\text{B}})^0 )$ terms that we neglect, one can replace each pole with the covariant form and also drop the ellipses. Projecting both sides to the $S$-wave, and substituting Eq.~\eqref{eq:FBdef}, we deduce Eq.~\eqref{eq:W_BS}.
 
\section{Ward-Takahashi identity for $\2+\Jc\to\2$ amplitudes}\label{app:WTI}

In this appendix, we demonstrate how Eq.~\eqref{eq:WTI_W} follows from the Ward-Takahashi identity. A consequence of current conservation, the Ward-Takahashi identity relates a given $n$-point Green function, coupled to an external conserved current, to the corresponding $(n-1)$-point Green function in which the current is omitted. Let $\Cc^{\mu}$ be a 5-point function coupling the conserved vector current, $\mathcal J^\mu$, to two neutral and two charged mesons. The Ward-Takahashi identity then reads
\begin{align}\label{eq:WTI_green}
q_{\mu}   \Cc^{\mu}(p',k';p,k) & = {\mathrm Q_0} \Big[   \Cc(p' + q,k' ; p,k) - \Cc(p',k';p-q,k)   \Big] \,,
\end{align}
where $q^{\mu} = (p'+k')^\mu - (p+k)^\mu =  P'^{\mu} - P^{\mu}$, with the second equality introducing notation for the total momenta of the outgoing and incoming two-meson states. We have also introduced $\Cc$ (with no index) as the four-point function without the current insertion. We further define $k$ and $k'$ as the initial- and final-state momenta of the neutral particles, respectively, and $p = P- k$ and $p' = P' - k'$ as the corresponding momenta for the particles carrying the charge, $ {\mathrm Q_0} $.

The $\2 + \Jc \to\2$ and $\2\to\2$ amplitudes, $\mathcal W^\mu$ and $\mathcal M$ respectively, are related to the Green functions by amputating the external meson propagators and placing them on the mass shell, \ie,
\begin{align}
\Cc_{\textrm{amp}} & \xrightarrow[\textrm{on-shell}]{} \Mc \,, \\ 
\Cc_{\textrm{amp}}^{\mu}  & \xrightarrow[\textrm{on-shell}]{} \Wc^{\mu},
\end{align}
where the amputated Green functions are defined as
\begin{equation}\label{eq:C_amp}
\Cc_{\textrm{amp}} (p',k';p,k) \equiv  (p'^2 - m^2) (k'^2 - m^2) (p^2 - m^2) (k^2 - m^2)  \,  \Cc (p',k';p,k)   \,,
\end{equation}
and the same with the $\mu$ index included on both sides.  Considering only the amputation at this stage and substituting Eq.~\eqref{eq:C_amp} into \eqref{eq:WTI_green}, we find
\begin{equation}\label{eq:WTI_amp}
q_{\mu}\Cc_{\textrm{amp}}^{\mu}(p',k';p,k) = {\mathrm Q_0} \bigg[   \Cc_{\textrm{amp}}(p',k';p+q,k) \frac{p^2 - m^2}{(p+q)^2 - m^2} - \frac{p'^2 - m^2}{(p'-q)^2 - m^2} \Cc_{\textrm{amp}}(p'-q,k';p,k)   \bigg] \,,
\end{equation}
where the ratios of amputation factors arise since the Ward-Takahashi identity changes the momenta carried by the mesons on the two sides of the equation.

In the limit where $p', k', p$ and $k$ go on shell, the numerators on the right hand side of Eq.~\eqref{eq:WTI_amp} vanish but the denominators do not, yielding the well-known Ward identity: $q_{\mu} \Wc^{\mu} = 0$. In addition, the long-range pieces that define the difference between $\mathcal W^\mu$ and $\Wc_{\df}^{\mu}$ [see Fig.~\ref{fig:2Jto2_W_amp}(b)] are proportional to $(P'+P)^\mu$ and therefore also vanish when contracted with $q_\mu$. (Equivalently they are proportional to the single-particle matrix element of $\mathcal J^\mu$ and must therefore also satisfy the Ward identity.) It follows that $\Wc_{\df}^{\mu}$ itself satisfies the identity: $q_{\mu} \Wc^{\mu}_{\df} = 0$.

Returning to the off-shell relation,  Eq.~\eqref{eq:WTI_amp}, we re-express all functions in terms of $P, P', k$ and $k'$ to write
\begin{multline}\label{eq:WTI_amp}
  {q_{\mu}\Cc_{\textrm{amp}}^{\mu}(P' - k',k';P-k,k)}  =\\    \mathrm Q_0 \bigg [  \Cc_{\textrm{amp}}(P' - k',k'; P' - k,k)  \frac{(P-k)^2 - m^2}{(P'-k)^2 - m^2} - \frac{(P' - k ')^2 - m^2}{(P-k')^2 - m^2} \Cc_{\textrm{amp}}(P - k',k';P - k,k)  \bigg ]    \,.
\end{multline}
Applying a $P'_\nu$ derivative on the left-hand side then gives
\begin{align}\label{eq:WTI_amp}
\frac{\partial}{\partial P'_{\nu}} [\text{LHS}] =  { \Cc_{\textrm{amp}}^{\nu}(P' - k',k';P-k,k)} +   q_{\mu}   \frac{\partial \, \Cc_{\textrm{amp}}^{\mu}(P' - k',k';P-k,k)} {{\partial P'_{\nu}} } \,,
\end{align}
and, applying the same to the right-hand side, one finds
\begin{align}\label{eq:WTI_amp}
\frac{\partial}{\partial P'_{\nu}} [\text{RHS}] &=  \mathrm Q_0 \frac{\partial  \, \Cc_{\textrm{amp}}(P' - k',k'; P' - k,k) }{\partial P'_\nu}    \frac{(P-k)^2 - m^2}{(P'-k)^2 - m^2}  \nn \\
& -   2  \mathrm Q_0(P' - k)^{\nu}   \Cc_{\textrm{amp}}(P - k',k'; P - k,k) \frac{(P-k)^2 - m^2}{[(P'-k)^2 - m^2]^2} \nn \\
&  - \frac{ 2 \mathrm Q_0 (P' - k)^{\nu}}{(P-k')^2 - m^2} \Cc_{\textrm{amp}}(P - k',k';P - k,k)   \,.
\end{align}

Next, before equating the two sides, we take the zero-momentum-transfer limit ($P' \to P$) and substitute 
\begin{equation}
w^{\mu}(P-k;P-k) = 2(P-k)^{\mu} {\mathrm Q_0} \,,
\end{equation} 
for the $\1+\Jc\to\1$ matrix element at zero momentum transfer. This then gives
\begin{align}
\Cc_{\textrm{amp}}^{\mu} (P - k',k';P,P-k) & = {\mathrm Q_0} \frac{\partial}{\partial P_{\mu}} \Cc_{\textrm{amp}}(P - k',k';P-k,k)   \nn \\
& +  i\Cc_{\textrm{amp}}(P - k',k';P-k,k) \, \frac{i}{(P-k)^2 - m^2} \, w^{\mu}(P-k;P-k)  \nn   \\
& + w^{\mu}(P-k';P-k') \, \frac{i}{(P-k')^2-m^2} \, i\Cc_{\textrm{amp}} (P-k',k';P-k,k) \,.
\end{align}
Setting $P = P'$ has greatly simplified the expressions, but care must be taken as the second and third terms on the right-hand side will diverge when we set $p', k', p$ and $k$ to their on shell values. Indeed these are the same divergences that appear in the difference between $\Wc^{\mu}$ and $\Wc_{\df}^{\mu}$, with the only subtlety that they were first defined in on-shell amplitudes at $P' - P \neq 0$. Fortunately, in the present case the distinction is unimportant because, when applied to the divergence-free amplitude, the zero-momentum-transfer and on-shell limits commute. We can thus move the second and third terms to the left-hand side and
take $p', k', p,k$ on shell to conclude
\begin{equation}
\Wc_{\df}^{\mu}(P,\bh{\k}_f^{\star}; P, \bh{\k}_i^{\star})  =    {\mathrm Q_0} \frac{\partial}{\partial P_{\mu} } \Mc(s,\bh{\k}_f^{\star},\bh{\k}_i^{\star})  \,.
\end{equation}
This remarkable result gives a clear interpretation to $\Wc_{\df}^{\mu}$ in the forward limit.

Finally, since the derivative is with respect to total momenta, we can easily project both sides to definite angular momentum. This leads to
\begin{equation}
\Wc_{\df, \ell' m' , \ell m}^{\mu}(P) = \delta_{\ell' \ell} \delta_{m' m } \, {\mathrm Q_0} \frac{\partial}{\partial P_{\mu} }\, \Mc_{\ell}(s) \,,
\end{equation}
as claimed in Eq.~\eqref{eq:WTI_W} for the special case of $S$-wave systems.

\section{Analytic continuation of finite-volume functions below threshold}\label{app:cont}
 
 In this section we give results for the analytic continuations of the $F$- and $G$-functions below threshold. Specifically we require results for $F(P,L)$, $G(P,L)$ and $G^{\mu = 0}(P,L)$, where we recall that a single momentum argument within $G$ indicates that the initial- and final-state four-momenta are equal. Each of these can be written in terms of a class of functions naturally extending those defined in Refs.~\cite{Luscher:1986pf,Luscher:1991n1,Rummukainen:1995vs,Kim:2005gf}:%
 \footnote{The $c_{JM}^{(n)}$ are proportional to the dimensionless functions denoted by $\Zc_{JM}^{(n)}$ in Ref.~\cite{Baroni:2018iau}
\begin{equation}
c_{JM}^{(n)}(P,L) = \frac{L^{2n-3}}{(2\pi)^{2n}} \left( \frac{2\pi}{L} \right)^{J} \Zc_{JM}^{(n)}(P,L) \,.
\end{equation}
For the analytic work presented here, the dimensionful versions prove slightly more convenient.}
\begin{equation}\label{eq:cJKM}
c_{JM}^{(n)}(P,L) =  
\bigg[  \frac{1}{L^{3}}  \SumInt_{\k} \bigg]
 \frac{\omega_\k^{\star}}{\omega_{\k}} \frac{\sqrt{4\pi} \, k^{\star\,J} Y_{JM}(\bh{\k}^{\star})}{(q^{\star\,2} - k^{\star\,2} + i\epsilon)^{n}}  \,.
\end{equation}

 The relations to the finite-volume functions that we require are then given by
 \begin{align}
\label{eq:Fc1_def}
F(P,L) & =   \frac{1}{2 E^{\star}}    c^{(1)}_{00}(P,L) \,, \\
G(P,L) & = \frac{1}{4E^{\star}} c^{(2)}_{00}(P,L) \,, \\
G^{\mu = 0}(P,L) & = -\frac{E}{4E^{\star\,3}} c_{00}^{(1)}(P,L) + \frac{ E }{8 E^{\star}} c_{00}^{(2)}(P,L) + \frac{1}{4\sqrt{3}} \frac{P_{z}}{E^{\star\,2}} c_{10}^{(2)}(P,L) \,,
\end{align}
where the last result also assumes that $\textbf P$ is parallel to the $\hat {\textbf z}$ axis.

For $P^2 < (2m)^2$, the summand of $c^{(n)}$ is a smooth function of $\textbf k^\star$ with a finite region of analyticity. As a result, the sum and integral must become exponentially close to each other, with the scale in the exponential given by the grid-spacing of the sum (set by $L$) and the size of the analytic domain (set by $4 m^2 - P^2$). To make this explicit, we apply the Poisson summation formula to $c_{JM}^{(n)}$, evaluated at a generic sub-threshold four-momentum, $P_{\kappa}$, satisfying $m^2 - P_{\kappa}^2/4 = \kappa^2 $. We find
\begin{align}
\label{eq:KSS_general}
c_{JM}^{(n)}(P_{\kappa},L)  
&=
 (-1)^{n}\sum_{\m\ne\0}\int \frac{\diff^3\k^{\star}}{(2\pi)^{3}}  \frac{\sqrt{4\pi} k^{\star\,J} Y_{JM}(\bh{\k}^{\star})}{(\kappa^{2} + k^{\star\,2} )^{n}}  e^{i L \m \cdot \k} \,,
\end{align}
where we have used the fact that the integration measure is a Lorentz invariant, $\diff^3 \k / \omega_{\k} = \diff^3 \k^{\star} / \omega_{\k}^{\star}$.

The kinematic variables in the CMF are related to the moving frame variables via standard Lorentz transformations,
\begin{align}
\k_{||}^{\star} & = \gamma(\k_{||} -  \omega_{\k} \,  \bs{\beta}) , \nn \\
\k_{\perp}^{\star} & = \k_{\perp} \,, \nn \\
\omega_{\k}^{\star} & = \gamma(\omega_{\k} - \bs{\beta} \cdot \k),
\end{align}
where $\k_{\perp} = \k - \k_{||}$, $\k_{||} = (\k \cdot \bh{\bs{\beta}}) \bh{\bs{\beta}}$, $\bs{\beta} = \P / E$ is the velocity, and $\gamma = E / E^{\star}$ the Lorentz factor. We can then write the phase factor in terms of the CMF momenta, 
\begin{equation}
\m \cdot \k = \m' \cdot \k^{\star} +   \frac{\omega_{\k}^{\star}}{E^{\star}} \, \m \cdot \P,
\end{equation}
with $\m'$ defined in Eq.~\eqref{eq:mp}.

With these relations in hand we can write the integrand solely in terms of $\k^{\star}$,
\begin{align}
c_{JM}^{(n)}(P_{\kappa},L) & = \frac{(-1)^{n}}{(2\pi)^{3}} \sum_{\m \ne \0} \int_{0}^{\infty} \diff k^{\star} \, \frac{(k^{\star})^{J + 2} }{(k^{\star\,2} + \kappa^2)^{n} }  e^{i L \omega_{\k}^{\star} \m\cdot\P / E^{\star}}  \int \diff \bh{\k}^{\star} \, \sqrt{4\pi} Y_{J M}(\bh{\k}^{\star}) e^{iL \m' \cdot \k^{\star}}.
\label{eq:CJMangularint}
\end{align}
Next, we evaluate the angular piece by introducing spherical Bessel functions and making use of the standard plane wave expansion,
\begin{align}
e^{iL \m' \cdot \k^{\star}} & = 4\pi \sum_{\ell = 0}^{\infty} i^{\ell} j_{\ell}(L \lvert\m'\rvert k^{\star}) \sum_{m_{\ell} = -\ell}^{\ell} Y_{\ell m_{\ell}}(\bh{\m}') Y_{\ell m_{\ell}}^{*}(\bh{\k}^{\star}) \,,
\end{align}
where $j_{\ell}(z)$ is the spherical Bessel function of the first kind. The angular integral in Eq.~\eqref{eq:CJMangularint} becomes
\begin{align}
\mathcal{I}_{JM}(L k^{\star}|\m'|,\m)  
&\equiv
\int \diff \bh{\k}^{\star} \, \sqrt{4\pi} Y_{J M}(\bh{\k}^{\star}) e^{iL \m' \cdot \k^{\star}} \,,\\
& = (4\pi)^{3/2}  i^{J} j_{J}(L \lvert \m' \rvert k^{\star})  Y_{JM}(\bh{\m}') \,.
\end{align}
For the cases considered here we require only
\begin{align}
\mathcal{I}_{00}(L k^{\star}|\m'|,\m) &=
\frac{4\pi\,\sin\left(L k^{\star}|\m'|\right)}{L   k^{\star}|\m'|},
\\
\mathcal{I}_{10}(L k^{\star}|\m'|,\m)  &=
i4\pi  \sqrt{3}
\left(\frac{\sin\left(L k^{\star}|\m'|\right)}{\left(L k^{\star}|\m'|\right)^2}
-\frac{\cos\left(L k^{\star}|\m'|\right)}{L k^{\star}|\m'| }
\right)\,
  \frac{  {\m}' \cdot   \P}{ \vert \m ' \vert \vert \P \vert} \,.
\end{align}

To evaluate the remaining integral over $k^\star$, we express the sinusoidal functions in $\mathcal{I}_{JM}$ in terms of exponentials and then divide the function $\mathcal{I}_{JM}$ into two terms, denoted $\mathcal{I}^{(+)}_{JM}$ and $\mathcal{I}^{(-)}_{JM}$: $\mathcal{I}^{(+)}_{JM}$ is defined by replacing the sinusoidal functions with the part of their exponential representation that decays as $k^\star \to i\infty$, e.g.~$\sin(x) \to e^{ix}/(2i)$, and $\mathcal{I}^{(-)}_{JM}$ is defined in the same way for the part that decays as $k^\star \to - i\infty$, e.g.~$\sin(x) \to - e^{-ix}/(2i)$.
This leads to a decomposition of $c_{JM}^{(n)}$ into two integrals 
\begin{align}
c_{JM}^{(n)}(P_{\kappa},L) & = \frac{(-1)^{n}}{(2\pi)^{3}} \sum_{x = \pm} \sum_{\m \ne \0} \int_{0}^{\infty} \diff k^{\star} \, \frac{(k^{\star})^{J + 2} }{(k^{\star\,2} + \kappa^2)^{n} }  e^{i L \omega_{\k}^{\star} \m\cdot\P / E^{\star}} \mathcal{I}^{(x)}_{JM}(L k^{\star}|\m'|,\m) \,.
\end{align}
Furthermore, one can show that the $\mathcal{I}^{(\pm)}_{JM}$ factors dominate the behavior at large, imaginary $k^\star$. It follows that the
$\mathcal{I}^{(+)}_{JM}$ ($\mathcal{I}^{(-)}_{JM}$) integral can be evaluated by closing the contour in the upper (lower) half of the complex plane. 
 
 In addition to the $k^\star=\pm i \kappa$ pole, the integrand has branch cuts starting at $k^\star=\pm i m$, associated with the square root in $\omega_\k$. These lead to exponential corrections of the order of $\mathcal{O}(e^{-mL})$ that are ignored throughout, i.p.~already in deriving the formalism considered in this work. Therefore these contributions should also dropped in the present evaluations. Keeping only the contribution from the $\kappa$ pole, we deduce
 \begin{align}
 \label{eq:cn100_asymp}
c_{00}^{(1)}(P_{\kappa},L) 
& 
 =
  -\sum_{\m\ne\0} 
 \,
 e^{iL \m \cdot \P / 2}
 \,
  \frac{ e^{-\kappa L \lvert\m'\rvert}}
  {4\pi L \lvert\m'\rvert}
\,,
 \\
 \label{eq:cn200_asymp}
c_{00}^{(2)}(P_{\kappa},L)  
& = 
 \sum_{\m \ne \0} e^{iL \m\cdot\P / 2} 
 \frac{e^{-\lvert\m'\rvert \kappa L}}{8\pi}
  \left( 
  \frac{1}{\kappa} 
  + 
  2\frac{i}{E^{\star 2} } \frac{\m \cdot \P}{\lvert\m'\rvert} \right) \,,
 \\
 \label{eq:cn200_asymp}
c_{10}^{(2)}(P_{\kappa},L)  
& = 
 \sqrt{3} \sum_{\m \ne \0} e^{iL \m\cdot\P / 2} 
 \frac{ e^{-\lvert\m'\rvert \kappa L}}{8\pi}
  \frac{  {\m}' \cdot   \P}{ \vert \m ' \vert \vert \P \vert}
  \,\left( 
i  
- 
2\frac{\m \cdot \P}{ E^{\star 2}  \, \lvert\m'\rvert }
\left(
\kappa
+\frac{1}{L\,\lvert\m'\rvert}
\right)
\right) \,.
\end{align}

When $\textbf P=0$ these expressions simplify significantly
\begin{align}
 \label{eq:cn100_asympP000}
c_{00}^{(1)}(P_{\kappa},L) 
& 
 =
  -\sum_{\m\ne\0} 
 \,
  \frac{ e^{-\kappa L \lvert\m \rvert}}
  {4\pi L \lvert\m \rvert}
\,,
 \\
 \label{eq:cn200_asympP000}
c_{00}^{(2)}(P_{\kappa},L)  
& = 
 \sum_{\m \ne \0}  
 \frac{e^{-\lvert\m \rvert \kappa L}}{8\pi \kappa}
  \,,
 \\
 \label{eq:cn200_asympP000}
c_{10}^{(2)}(P_{\kappa},L)  
& = 
0 \,.
\end{align}

\bibliography{bibi} %

\end{document}